\documentclass{aa} 

\usepackage{graphicx}
\usepackage{txfonts}
\usepackage{natbib}
\usepackage{float}

\begin{document}

\title{Ro-vibrational excitation of an organic molecule (HCN) in protoplanetary disks}

\author{Simon Bruderer\inst{\ref{inst:mpe}}, Daniel Harsono\inst{\ref{inst:leiden},\ref{inst:sron}}, and Ewine F. van Dishoeck\inst{\ref{inst:leiden},\ref{inst:mpe}}}

\institute{
Max-Planck-Institut f\"{u}r Extraterrestrische Physik, Gie\ss enbachstrasse 1, 85748 Garching, Germany\label{inst:mpe}
\and
Leiden Observatory, Leiden University, P.O. Box 9513, 2300 RA Leiden, The Netherlands\label{inst:leiden}
\and
SRON Netherlands Institute for Space Research, PO Box 800, 9700 AV, Groningen, The Netherlands\label{inst:sron}
}

\date{Accepted version -- \today}

\titlerunning{Excitation of HCN in protoplanetary disks}
\authorrunning{S. Bruderer et al.}

\offprints{Simon Bruderer,\\ \email{simonbruderer@gmail.com}}

\abstract
{Organic molecules are important constituents of protoplanetary disks. Their ro-vibrational lines observed in the near- and mid-infrared are commonly detected toward T Tauri disks. These lines are the only way to probe the chemistry in the inner few au where terrestrial planets form. To understand this chemistry, accurate molecular abundances have to be determined. This is complicated by excitation effects that include radiative pumping. Most analyses so far have made the assumption of local thermal equilibrium (LTE), which may not be fulfilled because of the high gas densities required to collisionally thermalize the vibrational levels of the molecules.} 
{The non-LTE excitation effects of hydrogen cyanide (HCN) are studied to evaluate \emph{(i)} how the abundance determination is affected by the LTE assumption, \emph{(ii)} whether the ro-vibrational excitation is dominated by collisions or radiative pumping, and \emph{(iii)} which regions of protoplanetary disks are traced by certain vibrational bands.} 
{Starting from estimates for the collisional rate coefficients of HCN, non-LTE slab models of the HCN emission were calculated to study the importance of different excitation mechanisms. Using a new radiative transfer model, the HCN emission from a full two-dimensional disk was then modeled to study the effect of the non-LTE excitation, together with the line formation. We ran models tailored to the T Tauri disk AS 205 (N) where HCN lines in both the 3 $\mu$m and 14 $\mu$m bands have been observed by VLT-CRIRES and the \emph{Spitzer Space Telescope}.} 
{Reproducing the observed 3 $\mu$m / 14 $\mu$m flux ratios requires very high densities and kinetic temperatures ($n >10^{14}$ cm$^{-3}$ and $T>750$ K), if only collisional excitation is accounted for. Radiative pumping can, however, excite the lines easily out to considerable radii $\sim 10$ au. Consequently, abundances derived from LTE and non-LTE models do not differ by more than a factor of about 3. Models with both a strongly enhanced abundance within $\sim 1$ au (jump abundance) and constant abundance can reproduce the current observations, but future observations with the MIRI instrument on JWST and METIS on the E-ELT can easily distinguish between the scenarios and test chemical models. Depending on the scenario, ALMA can detect rotational lines within vibrationally excited levels.} 
{Pumping by the continuum radiation field can bring HCN close enough to the LTE so that no big deviations in derived abundances are introduced with the LTE assumption, but the line profiles are substantially altered. In non-LTE models, accounting for collisional and radiative excitation, the emitting region can be much larger than in LTE models. Because HCN can be radiatively pumped to considerable radii, deriving a small emitting region from observations can thus point to the chemical abundance structure (e.g., jump abundance). Owing to their level structure, CO$_2$ and C$_2$H$_2$ are expected to act in a similar way, facilitating studies of the warm inner disk chemistry.} 

\maketitle

%
%

\section{Introduction} \label{sec:intro}

Organic molecules play an important role in the chemistry of planet-forming zones. These molecules are the initial step toward forming the prebiotic molecules that ultimately lead to life. They are also important tracers of the physical and chemical conditions in protoplanetary disks (\citealt{Bergin07b,Herbst09,Pontoppidan14b} for reviews). A surprising finding of the \emph{Spitzer Space Telescope} was the rich mid-infrared (mid-IR) spectrum from the innermost few au containing organic molecules (\citealt{Lahuis06,Salyk08,Carr08}). Different disk radii can be probed by combining submillimeter to near-IR wavelengths, since these lines span a wide range of level energies. New facilities such as the \emph{Atacama Large Millimeter/submillimeter Array} (ALMA), the \emph{Very Large Telescope} (VLT), or the future \emph{James Webb Space Telescope} (JWST) and \emph{E-European Extremely Large Telescope} (E-ELT), pave the way for such multiwavelength observations at high spatial resolution and good sensitivity. To make progress in understanding the chemical structure of protoplanetary disks and the chemical history of planetary atmospheres and comets, accurate abundances have to be derived. This is complicated by excitation and radiative transfer effects. In this work, we study the non-LTE vibrational excitation and radiative transfer effects of HCN. 

Hydrogen cyanide (HCN) is a simple organic molecule, but an important reactant for producing more complex nitrogen-bearing molecules, building blocks of amino acids such as adenine (\citealt{Agundez08,Walsh14,Bast13}). Together with acetylene (C$_2$H$_2$) and methane (CH$_4$), HCN is one of the bottlenecks in producing more complex species. Owing to the high activation energies of the formation reactions, these molecules are strongly enhanced in the innermost 1 au of a typical T Tauri disk (\citealt{Agundez08,Bast13}). For example, ${\rm CN} + {\rm H}_2 + \Delta E \rightarrow {\rm HCN} + {\rm H}$ has an activation barrier of $\Delta E = 820$ K. Because of the enhanced HCN abundance in the innermost region of protoplanetary disks, this molecule may be one of the best tracers of the conditions in terrestrial planet-forming regions.

Trends of HCN emission relative to other molecules have been found observationally. An increase in the mid-IR HCN/H$_2$O ratio with disk mass measured by the submillimeter continuum is found by \cite{Najita13} (see also \citealt{Najita11}) and explained by a larger number of non-migrating planetesimals in more massive disks locking up water ice beyond the snow line. However, toward very low mass stars, this trend does not hold (\citealt{Pascucci13}). The HCN/C$_2$H$_2$ mid-IR emission ratio was found to be significantly higher around Sun-like stars compared to brown dwarfs which was interpreted by a different physical and chemical evolution of the protoplanetary disks (\citealt{Pascucci09}). So far, these trends have only been discussed in the context of simple LTE slab models. However, to answer the question if trends are the consequence of excitation or abundance structure, a more complete modeling accounting for non-LTE excitation effects needs to be carried out.

HCN has a dipole moment and can be observed through pure rotational lines at submillimeter wavelengths and ro-vibrational transitions in the infrared. Besides CO and H$_2$O, HCN is currently one of the few molecules with such multiwavelength detections towards protoplanetary disks. It is detected at submillimeter wavelengths (\citealt{Dutrey97,Kastner97,vanZadelhoff01,Thi04,Oberg10,Oberg11,Chapillon12}), in the mid-IR (\citealt{Carr08,Pontoppidan10,Salyk11}), and in the near-IR (\citealt{Gibb07,Doppmann08,Mandell12}). Towards massive young stellar objects rotational lines of HCN within vibrationally excited levels have also been observed (e.g., \citealt{Boonman01,Benz07,Veach13}). Vibrational lines\footnote{Note that different definitions of the vibrational modes are used. Here, $\nu_1$ refers to the H-C stretching mode, $\nu_2$ to the bending mode, and $\nu_3$ to the C-N stretching mode, following the definition by the NIST.} from the $\nu_2$ bending mode at 14 $\mu$m have been observed spectrally unresolved by \emph{Spitzer} (\citealt{Lahuis06}). The $\nu_1$ H-C stretching mode at 3 $\mu$m is observed with the VLT \emph{CRyogenic high-resolution InfraRed Echelle Spectrograph} (CRIRES) and NIRSPEC on Keck. In the near future, the \emph{VLT Imager and Spectrometer for mid Infrared II} (VISIR-II) and the \emph{Texas Echelon Cross Echelle Spectrograph} (TEXES) on Gemini North will be able to spectrally resolve the 14 $\mu$m lines. With the \emph{Mid-Infrared Instrument} (MIRI) and the \emph{Near-Infrared Spectrograph} (NIRSpec) on the JWST and the \emph{Mid-infrared E-ELT Imager and Spectrograph} (METIS), it will be possible to study the entire $3-14$ $\mu$m wavelength range at much higher sensitivity and spectral resolution also allowing observation of the $2 \nu_2 - \nu_2$ mode at 7 $\mu$m.

The analysis of HCN vibrational lines has so far mostly relied on the LTE assumption. This means that the HCN level population is driven to thermalization, usually assumed to be by collisions with the main constituent of the molecular gas, H$_2$. However, the so-called \emph{critical density} $n_{\rm crit}$, needed to provide enough collisions for thermalization, is for vibrational transitions orders of magnitude larger than for pure rotational lines ($n_{\rm crit} \sim 10^8$ cm$^{-3}$ for HCN $J=4-3$). Towards massive protostars, \cite{Boonman03c} thus estimate based on a simple model, that HCN is more likely pumped by infrared continuum radiation of the warm dust than collisionally excited. This is similar to the vibrational bands HCN in comets which are pumped radiatively by solar radiation (e.g., \citealt{Paganini10}). In protoplanetary disks, the critical densities of vibrational transitions can be reached. However, the dust optical depths towards such dense regions is likely large and shields the molecular emission. A comprehensive model of the HCN emission from protoplanetary disks should thus account for radiative pumping and collisional excitation effects as well as the line formation process like shielding of line emission by dust. So far, only a few studies on the vibrational excitation of molecules in protoplanetary disks have been carried out, like \cite{Meijerink09} on water and \cite{Blake04,Thi13b} on CO.

In this work, the non-LTE emission of HCN from protoplanetary disks is modeled. We derive an approximate set of ro-vibrational collision rate coefficients, based on measured values and calculate the non-LTE emission from a simple slab to study the relative importance of different processes. We next model the HCN emission of the protoplanetary disk AS 205 (N), where both the 14 $\mu$m band has been observed by \emph{Spitzer} (\citealt{Salyk11}) and the 3 $\mu$m lines by VLT-CRIRES (\citealt{Mandell12}). While this work concentrates on this particular disk as an example, a parameter study is carried out to find the general trends. The main questions to be answered by our work are: How much does the abundance determination change if non-LTE effects are included? How are different vibrational bands of HCN excited (radiative or collisional excitation)? What do different HCN lines tell about the radial and vertical abundance structure?

The paper is organized as follows: In Sect. \ref{sec:hcn_excite} the molecular structure is summarized and slab models run to study the relative importance of collisional and radiative excitation. Sect. \ref{sec:hcn_disk} presents non-LTE models of a protoplanetary disk and a study of the dependence of the emission on various parameters like the HCN abundance structure, the gas-to-dust ratio or the dust opacity. The results are discussed in Sect. \ref{sec:discussion} followed by a conclusions section (Sect. \ref{sec:conclusions}). 

%
%

\section{The ro-vibrational excitation of HCN} \label{sec:hcn_excite}

To study the relative importance of different excitation mechanisms independent of geometry, we first discuss simple slab models at constant physical conditions (density, kinetic temperature and continuum radiation field). The steady state level population is then given by the rate equations (e.g., \citealt{vdTak07}),
\begin{equation} \label{eq:rates}
\frac{dn_i}{dt} = \sum_{i \neq j} n_j P_{ji} - n_i \sum_{i \neq j} P_{ij} = 0 \ ,
\end{equation}
where $n_i$ (cm$^{-3}$) is the population of a level designated with an index $i$ and $P_{ij}$ (s$^{-1}$) is the rate coefficient for the transition between levels $i$ and $j$. The rates of different collisional and radiative processes enter through 
\begin{equation}
P_{ij} = \left\{
\begin{array}{ll}
A_{ij}+B_{ij} \langle J_{ij} \rangle + K_{ij} n & (E_i > E_j) \\
B_{ij} \langle J_{ij} \rangle + K_{ij} n & (E_i < E_j) \ , \\
\end{array}
\right.
\end{equation}
with the Einstein-$A_{ij}$ and $B_{ij}$ coefficients for spontaneous and induced emission between levels $i$ and $j$, the local radiation field $\langle J_{ij} \rangle$ averaged over angles and local line profile function, the collisional rate coefficient $K_{ij}$, and the collisional partner density $n$. Spontaneous decay from level $i$ to level $j$ is only possible if the level energy $E_j < E_i$. The level population controls the emission of line photons and the line opacity. Thus, through the radiation field $\langle J_{ij} \rangle$ the level population enters Eq. \ref{eq:rates} complicating the problem considerably. Once a self-consistent solution of level population and radiation field is found, the emitted emission is obtained from the solution of the radiative transfer equation.

The slab models presented in this section are calculated using the RADEX code (\citealt{vdTak07}). This code approximates the contribution of the line photons to the radiation field by an analytical function of the level population, using the so-called escape probability formalism (e.g., \citealt{Sobolev1960}). Initial exploration with a limited model was performed by \cite{Bast13b}.

\subsection{A model molecule for HCN} \label{sec:hcn_molec}

\begin{figure}[htb!]
\center
\includegraphics[width=1.0\hsize]{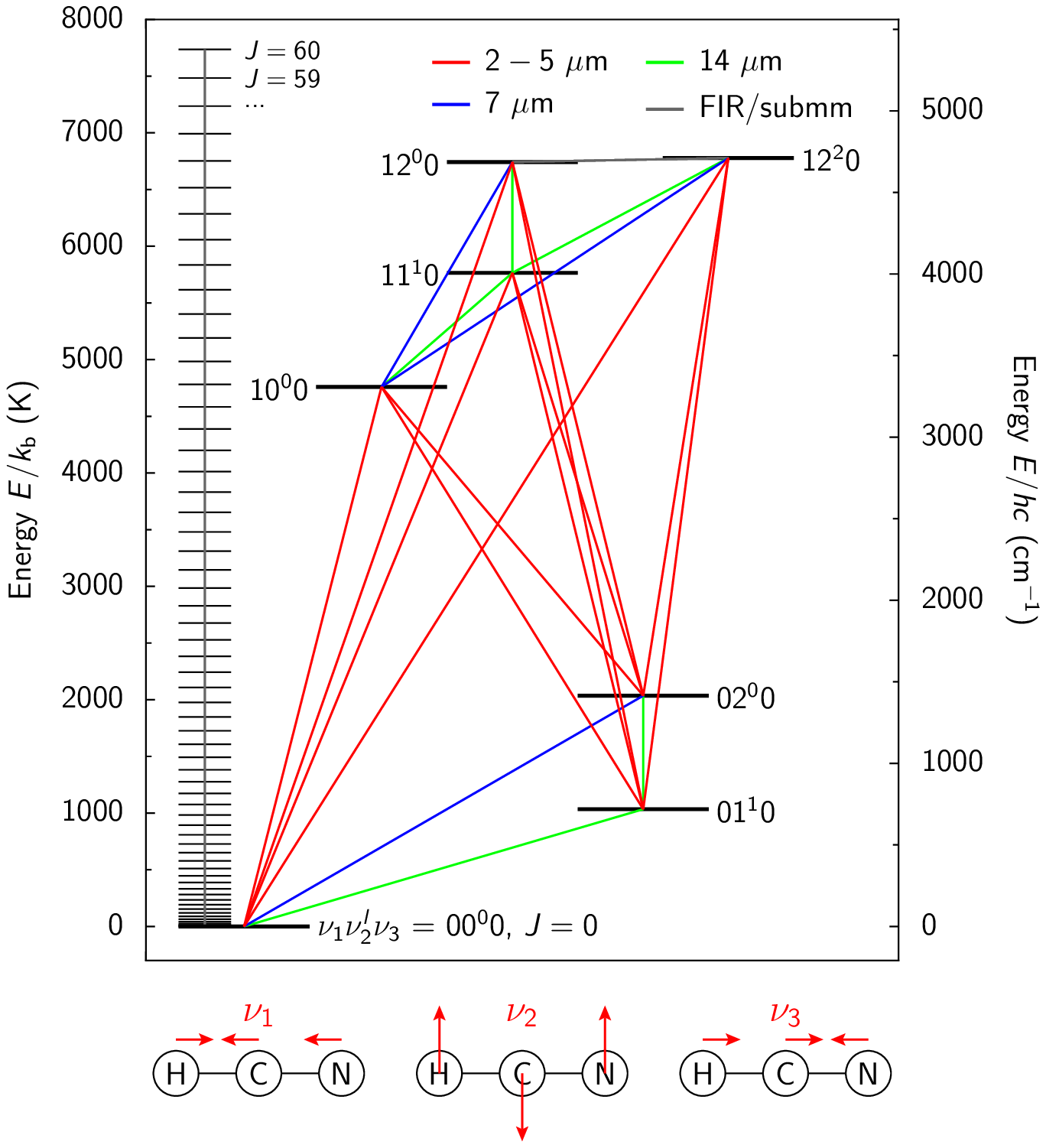}
\caption{HCN level structure considered in this work. Red lines denote vibrational bands at wavelengths of $2-5$ $\mu$m, while blue and green lines indicate bands at 7 $\mu$m and 14 $\mu$m, respectively. The vibrational modes $\nu_1$, $\nu_2$, and $\nu_3$ are shown below the level diagram. For clarity, the rotational ladder is only shown for the vibrational ground state.}
\label{fig:plot_lev}
\end{figure}

For the excitation calculation a model molecule of HCN needs to be constructed. The same 7 vibrational levels are included as \cite{GonzalezAlfonso99} and \cite{Cernicharo99} used to model the vibrational emission of radiatively pumped HCN towards AGB stars. Figure \ref{fig:plot_lev} shows the included energy levels and vibrational band wavelengths. We include $l$-type splitting of the levels in the $\nu_2$ bending mode with simultaneous rotation along the internuclear axis ($l>0$). Including 3 levels with $l$-type splitting, there is thus a total number of 10 vibrational levels, each with 60 rotational levels. The effect of other vibrational levels is not expected to become important, as they are not radiatively connected to the ground state (\citealt{GonzalezAlfonso99}). In particular, the $\nu_3$ modes have very low Einstein-A coefficients and are not included. Hyperfine splitting is not included, as our observations do not spectrally resolve individual hyperfine components. The electronic structure of HCN is quite complicated, due to intramolecular isomerization HCN - HNC above $\sim 2.1$ eV (e.g., \citealt{Barger03}) but this does not affect the levels considered here. The photoabsorption bands in the UV are all dissociative (\citealt{Lee84}) and we thus do not account for UV pumping or fluorescence. Chemical pumping has been suggested to explain a submillimeter maser within the $04^00$ band observed towards \mbox{IRC $+$10216}, but is unlikely to affect lower lying vibrational bands (\citealt{Schilke00}).

The level energies and radiative transition probabilities (Einstein-$A$ coefficients) are taken from \cite{Barber14}. To model H$^{13}$CN, energies and transition probabilities by \cite{Harris08} are used. A total of 602 levels and 4622 radiative transitions are included. No collisional rate coefficients for the vibrational levels of HCN are available. As for CO or water approximations are needed here (\citealt{Faure08,Thi13b}). Based on measured H$_2$ - HCN collisional rate coefficients for the total $1 1 0 \rightarrow 0 0 0$ and $1 0 0 \rightarrow 0 0 0$ bands (\citealt{Smith91}), a full set of collisional rate coefficients is constructed by combining these measured vibrational rates with pure rotational collision rates by \cite{Dumouchel08}. Our procedure follows the approach of \cite{Faure08} and \cite{Thi13b} and is described in detail in Appendix \ref{sec:app_coll}. The main assumption is that vibrational and rotational collisional rates can be decoupled. An assessment of the collisional rate uncertainty is difficult, but they are probably within an order of magnitude as estimated for the water rates by \cite{Faure08}. We only account for H$_2$-HCN collisions, as the critical densities of vibrational transitions are very high ($>10^{10}$ cm$^{-3}$) and gas is purely molecular in protoplanetary disks even at much lower densities.

To test the completeness of our model molecule, we have calculated the AS 205 (N) models presented in Table \ref{tab:flx_constjump} also with a reduced model molecule including only the vibrational ground state and the $\nu_1$ and $\nu_2$ modes. This reduced model molecule yields fluxes within 20 \% compared to the larger model molecule. The small deviation is understood by the fact that most of the molecules remain in the vibrational ground state throughout the disk and radiative pumping through a vibrational band is followed by decay through the same band.

\subsection{The HCN spectrum} \label{sec:hcn_spectrum}

\begin{figure*}[htb!]
\center
\includegraphics[width=1.0\hsize]{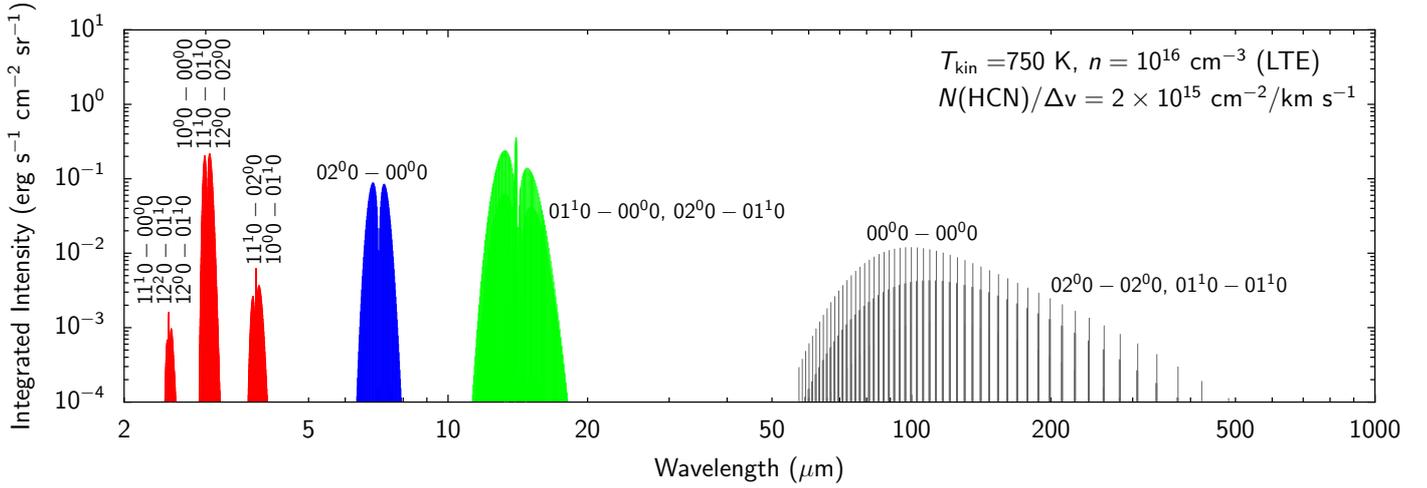}
\caption{HCN integrated intensity for a density of 10$^{16}$ cm$^{-3}$, a kinetic temperature of 750 K and a HCN column density of $2 \times 10^{15}$ cm$^{-2}$ at a line width of $\Delta {\rm v}=1$ km s$^{-1}$. For this density, the lines are close to the local thermal equilibrium (LTE). The same color scheme as in Figure \ref{fig:plot_lev} is used. Only the strongest bands are labeled.}
\label{fig:plot_fullspec}
\end{figure*}

A full spectrum of HCN from 2 $\mu$m to the pure rotational submillimeter lines from a slab model is shown in Figure \ref{fig:plot_fullspec}. We chose a kinetic temperature of 750 K and a column density of $2 \times 10^{15}$ cm$^{-2}$, in agreement with typical values found by \citet{Salyk11}. A Gaussian line profile with a FWHM line width of $\Delta {\rm v}=1$ km s$^{-1}$ is adopted\footnote{This intrinsic line width should not be confused with the observed, much broader line, resulting from the disk rotation.}. Except for some pure rotational lines at submillimeter wavelengths, the HCN lines remain optically thin. The density for this spectrum is chosen high enough to thermalize the HCN levels so that vibrational levels with high critical densities are also populated.

The strongest bands of this spectrum are at $2.9 - 3.2$ $\mu$m in the near-IR, at $6-8$ $\mu$m and $11-18$ $\mu$m in the mid-IR and at wavelengths $>50$ $\mu$m in the submillimeter. The near-IR feature consists of the $10^00-00^00$ band, which is a factor of 10 stronger than $12^00-02^00$ and 100 times stronger than $11^10-01^10$. These bands do not have a Q-branch ($\Delta J=0$), since $\Delta l=0$ ($\Sigma^+ \rightarrow \Sigma^+$). This is also the case in the $6-8$ $\mu$m feature centered at 7.08 $\mu$m. The 11-18 $\mu$m feature centered at 13.95 $\mu$m has $\Delta l=1$ ($\Pi \rightarrow \Sigma^+$) and thus a strong Q-branch. Pure rotational lines within the three vibrational ladders with the lowest energies ($00^00$, $01^10$, and $02^00$) contribute to the submillimeter lines. The $l$-type splitting shifts the ground states of the vibrational ladders by less than 0.5 GHz and thus cannot be seen in this spectrum (see \citealt{Thorwirth03}).

\subsection{Dependence on kinetic temperature, density and radiation field} \label{sec:depparam}

\begin{figure*}[htb!]
\center
\includegraphics[width=0.8\hsize]{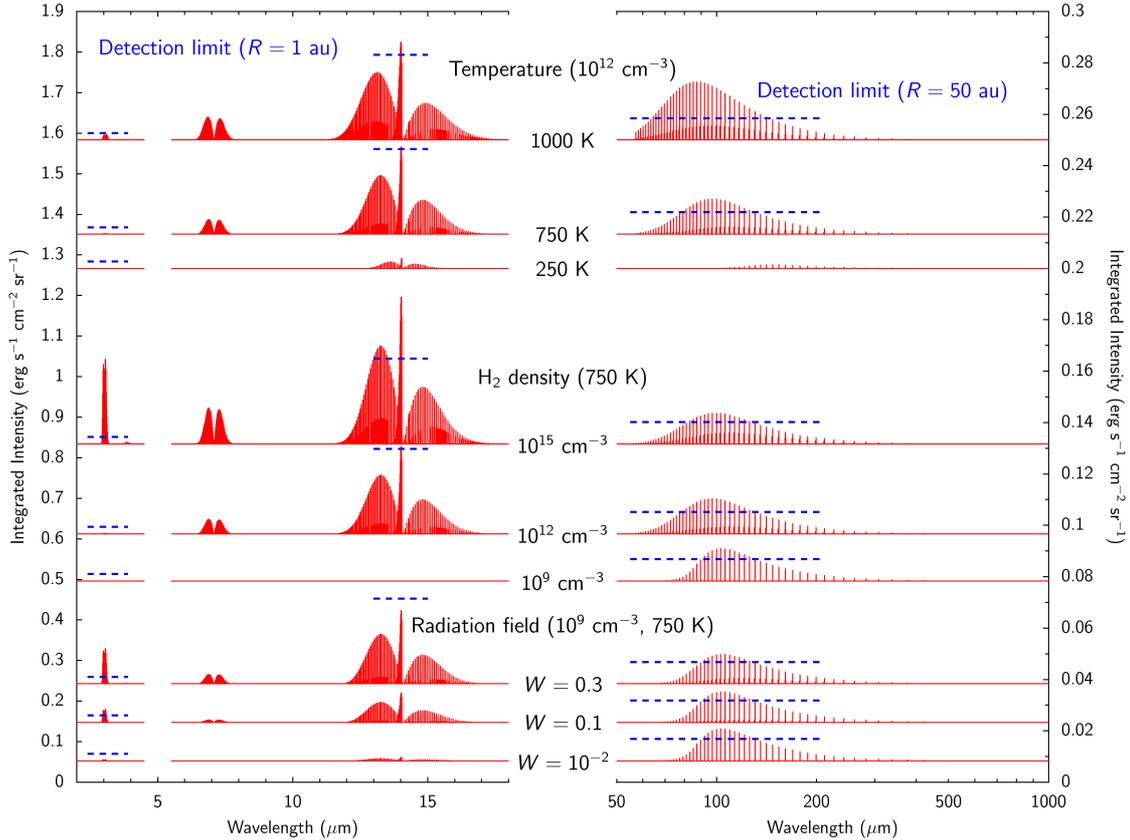}
\caption{HCN spectrum depending on kinetic temperature, density and radiation field for a slab model. The HCN column density is fixed to $2 \times 10^{15}$ cm$^{-2}$ and the line width to $\Delta {\rm v}=1$ km s$^{-1}$. The intensity scales of the IR (2-18 $\mu$m) and far-IR/submillimeter portions of the spectrum are different. The blue dashed line shows the typical detection limit of CRIRES and \emph{Spitzer} assuming an emitting region with radius $R=1$ au. For \emph{Herschel}-PACS, the detection limit within 50 au is shown.}
\label{fig:plot_paramspec}
\end{figure*}

The HCN excitation and line emission depends on kinetic temperature, collisional partner density and ambient radiation field. In Figure \ref{fig:plot_paramspec}, full HCN spectra with varied parameters are compared. To study the temperature and collisional partner density, no external radiation field is included. The dependence on an external (continuum) radiation field is then studied by including a diluted black-body $\langle J_{\nu,{\rm ext}} \rangle = W B_\nu(T_{\rm rad})$ with $T_{\rm rad}=750$ K. This $T_{\rm rad}$ has been adopted to reproduce the observed flux ratios towards AS 205 (N) (Sect. \ref{sec:comp_obs_slab}). The parameter range explored here approximately corresponds to the inner protoplanetary disk (Sect. \ref{sec:hcn_disk}). Typical detection limits for emission from the innermost region of a disk ($R=1$ au at 100 pc) are shown for VLT-CRIRES and \emph{Spitzer} (\citealt{Salyk11,Mandell12}). The detection limits scale as $R^2$ and four times weaker lines can be detected if $R=2$ au. For the far-IR lines within the wavelength range of \emph{Herschel}-PACS ($55-210$ $\mu$m) the detection limit within $R=50$ au at 100 pc is shown, corresponding to the emitting size found with PACS for other molecules (OH, H$_2$O, CO, see \citealt{Fedele13a,Meeus13}).

Increasing the kinetic temperature from 250 to 1000 K yields higher intensities in almost all lines, most noticeably in the near-IR and high-$J$ pure rotational transitions at submillimeter and far-IR wavelengths. This is because the shorter wavelength lines involve higher energy levels and the pure rotational ladder also extends up to several 1000 K for $J=60$ (Figure \ref{fig:plot_lev}). In the case of optically thin and thermalized lines, the emission is approximately $I \propto \exp(-E/k T)$ for an upper level energy $E$. At densities of $10^{12}$ cm$^{-3}$ a temperature close to 1000 K is necessary to excite the IR lines sufficiently for a detection from the innermost 1 au of a protoplanetary disk.

Higher collisional partner densities help exciting the lines, bringing them close to thermalization. Levels are thermalized above the so-called critical density, defined as $n_{\rm crit}=A_{ij}/K_{ij}$ for a two-level system. Below the critical density, optically thin emission is $I \propto n/n_{\rm crit}$. For low-$J$ pure rotational lines, critical densities are on the order of $10^8$ cm$^{-3}$, e.g. for $J=4 \rightarrow 3$. Thus, the pure rotational lines do not change considerably between densities of $10^9$ cm$^{-3}$ and $10^{15}$ cm$^{-3}$, because these lines are already thermalized. On the other hand, the 14 $\mu$m lines have critical densities on the order of a few times $10^{12}$ cm$^{-3}$ and the 3 $\mu$m lines on the order of $10^{15}$ cm$^{-3}$. Thus, these bands only show considerable emission at very high densities in the context of protoplanetary disks.

Radiative pumping by a diluted black-body dominates over collisional excitation, if (e.g., \citealt{Tielens05}),
\begin{equation}
\frac{W}{\exp\left( h \nu / k T_{\rm rad} \right) -1} \gtrsim \frac{n}{n_{\rm crit}} \ .
\end{equation}
For the assumed $T_{\rm rad}=750$ K and a density of $10^9$ cm$^{-3}$, radiation already dominates over collisions for $W \gtrsim 10^{-3}$ (14 $\mu$m) and $W \gtrsim 6 \times 10^{-4}$ (3 $\mu$m). However a stronger radiation field is necessary to produce detectable line emission in the near/mid-IR (Figure \ref{fig:plot_paramspec}). 

\subsection{Comparison with observations} \label{sec:comp_obs_slab}

HCN lines both at 3 $\mu$m and 14 $\mu$m have been detected towards the T Tauri disk AS 205 (N) (northern component). Using the near-IR high-resolution ($R = \lambda / \Delta \lambda \sim 10^5$) spectrometer CRIRES on VLT, \cite{Mandell12} detected several HCN lines in the 3 $\mu$m band. The $10^00$-$00^00$ R(6) line has a flux of $4.7 \times 10^{-15}$ erg cm$^{-2}$ s$^{-1}$ at a calibration uncertainty of $\sim 30$ \%. The lines have a low line-to-continuum ratio, thus the detection with CRIRES is very challenging. Not much information can be extracted from the line profile or spectro-astrometry and we will here focus on the total line flux. With the infrared spectrograph IRS onboard the \emph{Spitzer Space Telescope}, \cite{Salyk11} found a total flux of $(2.2 \pm 0.2) \times 10^{-13}$ erg cm$^{-2}$ s$^{-1}$ for the spectrally unresolved Q-branch at $13.7-14.1$ $\mu$m.

\begin{figure*}[htb!]
\center
\sidecaption
\includegraphics[width=0.75\hsize]{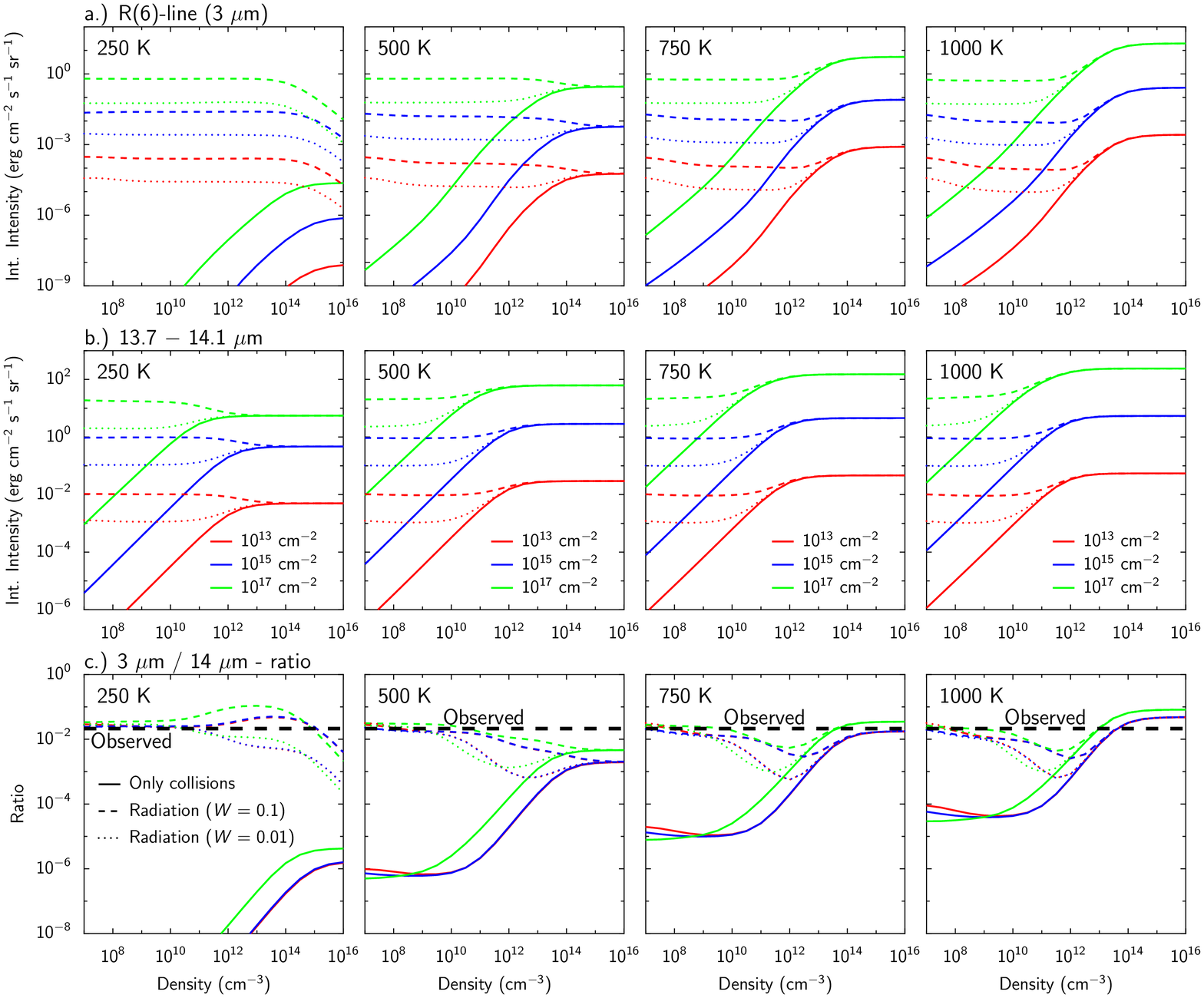}
\caption{Results of slab models. \textbf{a)} Integrated intensity of the R(6) $10^00-00^00$ line depending on the density for different kinetic temperatures and column densities. Solid lines show models that only consider collisions, while dashed and dotted lines represent models with a pumping radiation field of a 750 K blackbody diluted by $W=0.1$ (dashed line) or $W=0.01$ (dotted line). Red, blue, and green lines show results for column densities of $10^{13}$ cm$^{-2}$, $10^{15}$ cm$^{-2}$, and $10^{17}$ cm$^{-2}$, respectively. \textbf{b)} Total integrated intensity of all HCN lines between 13.7 and 14.1 $\mu$m for the same parameters as panel a). \textbf{c)} Ratio of panels a) and b) compared to the observed ratio towards AS 205 (N).}
\label{fig:plot_3ratio14}
\end{figure*}

For a first comparison of the non-LTE models with the observations, it is assumed that the 3 and 14 $\mu$m lines emerge from the same region of the disk and we will thus restrict our comparison to line ratios. This is motivated by the fact, that the size of the emitting region is difficult to constrain from current observations and often the assumption is made that the emitting size of HCN is that same as for water (e.g., \citealt{Salyk11}). We run RADEX models for densities between $10^7$ and $10^{16}$ cm$^{-3}$. The kinetic temperature is fixed to 250, 500, 750, or 1000 K and the column densities to $10^{13}$, $10^{15}$, or $10^{17}$ cm$^{-2}$ at an intrinsic line width $\Delta {\rm v}=1$ km s$^{-1}$. These column densities cover the range of values derived from observations (\citealt{Salyk11,Carr11}). Models with collisional excitation only or with additional radiation field from a 750 K blackbody diluted by $W=0.1$ or $W=0.01$ are explored.

Figure \ref{fig:plot_3ratio14} shows the integrated intensity\footnote{Scaled to flux in erg s$^{-1}$ cm$^{-2}$ by multiplying with the solid angle $2.35 \times 10^{-11} (R\, {\rm [au]}/D\, {\rm [pc]})^2$ sr, where $R$ is the radius of the emitting region in au and $D$ the distance to the object in parsec.} of the R(6) $10^00-00^00$ 3 $\mu$m line and the total $13.7-14.1$ $\mu$m emission and compares their line ratios with observations. For the models without radiation field, the intensity of both 3 $\mu$m and 14 $\mu$m lines increases linearly with density up to about their respective critical densities when they thermalize. Models with a pumping radiation field yield considerable emission even for low densities. For these models, the intensity increases with density only for the warm models (750 and 1000 K) since collisions thermalize the lines to a higher temperature than the radiation field. Thus, collisions may also decrease the line intensity. The increase of line intensity with radiation field ($W$) is about linear at low densities much below the critical densities of the transitions. In the limit of low densities, the radiation field thermalizes the level population to the temperature $T_{\rm rad}$. In the limit of high densities, collisions thermalize the level population to the kinetic temperature. Thus, in this example with $T_{\rm rad}=750$ K, the line ratios are driven to the same value ($\sim 2 \times 10^{-2}$) at low densities as they are at high densities in the panel for the kinetic temperature of 750 K. A lower/higher $T_{\rm rad}$ would decrease/increase the line ratio. For high column densities ($10^{17}$ cm$^{-2}$), the 14 $\mu$m lines get optically thick first, and the line ratio is increased compared to the optically thin case for the same kinetic temperature.

The comparison of the modeled and observed line ratios shows that pure collisional excitation requires very high densities $>10^{13}$ cm$^{-2}$ at kinetic temperatures $>750$ K. Meanwhile, a diluted radiation field with $T_{\rm rad}=750$ K can easily reproduce the observed line ratio, as long as the density is low enough to not affect the 14 $\mu$m lines which have a lower critical density.

%
%

\section{HCN emission from protoplanetary disks} \label{sec:hcn_disk}

More insight into the excitation and line formation process can be gained through models of protoplanetary disks including a realistic geometry. Such models have a range of physical conditions and not a single temperature and density as the slab models presented before. Using full disk models the size and location of the emitting regions can be constrained and effects of continuum shielding or radial/vertical variations in the abundance structure accounted for. This is not a trivial issue to calculate, since the pumping IR continuum that is mainly formed in the inner hot region, needs to be calculated at every position of the disk. Also, the molecular excitation needs to be calculated on a high resolution grid. In this section, we model the R(6) $10^00-00^00$ 3 $\mu$m line and 13.7-14.1 $\mu$m emission from AS 205 (N).

The model and its parameters are first introduced, the excitation of one particular model is presented and the effects of various parameters then studied. We emphasize that given the uncertainties in various input parameters (e.g., physical structure, dust opacities, \ldots), it is not the goal of this work to yield a perfect fit to the observed data, but to study the relative importance of different processes and parameters, which are also of relevance for disks other than AS 205 (N).

\subsection{Model setup}

Part of the physical-chemical model DALI (\textbf{D}ust \textbf{a}nd \textbf{Li}nes) is used in this work. Details of the model and benchmark tests are reported in \cite{Bruderer12} and \cite{Bruderer13}. Starting from a gas and dust density structure, the Monte-Carlo radiative transfer module calculates the dust temperature and the local continuum radiation field at each position of the disk. Then, the excitation module yields the HCN level population at all positions. The local radiation field is obtained using an escape probability approach (Appendix A.2 in \citealt{Bruderer13}) accounting for both line and continuum contributions. For the continuum pumping, the radiation field obtained by the continuum radiative transfer module, shielded by the line opacity, is used. Finally, synthetic spectra are derived with the raytracing module, which solves the radiative transfer equation along rays through the disk. The approach used here is akin to \cite{Meijerink09} and \cite{Thi13b} used for H$_2$O and CO, respectively. However, unlike \cite{Thi13b} a parameterized abundance structure is used instead of chemical network results. This allows us to study the excitation effects independently of uncertainties in the HCN chemistry. The assumption of $T_{\rm gas}=T_{\rm dust}$ will not affect the results, as HCN has very high critical densities and $T_{\rm gas} \sim T_{\rm dust}$ in regions where collisions are important.

The main input parameters for the AS 205 (N) model are taken from \cite{Andrews09}. They have simultaneously fitted SED and submillimeter images using a parameterized physical structure with surface density profile 
\begin{equation}
\Sigma(R) = \Sigma_{\rm c} \left( \frac{R}{R_{\rm c}} \right)^{-\gamma} \exp\left[-\left( \frac{R}{R_{\rm c}} \right)^{2-\gamma} \right] 
\end{equation}
and vertical distribution 
\begin{equation}
\rho(R,\theta) = \frac{\Sigma(R)}{\sqrt{2\pi} R \, h(R)} \exp\left[ -\frac{1}{2} \left( \frac{\pi/2 - \theta}{h(R)} \right)^2 \right] \ ,
\end{equation}
with the scale height angle $h(R) = h_{\rm c} ( R/R_{\rm c} )^\psi$. The values of the parameters and other basic properties adopted are summarized in Table \ref{tab:basic_as205}. 

\begin{table}[tbh]
\caption{Adopted basic data of AS 205 (N).}
\label{tab:basic_as205}
\centering
\begin{tabular}{lcc}
\hline\hline
Parameter & &  Value  \\
\hline
Star & & \\
Mass                    & $M_*$ [$M_\odot$]          & 1.0\tablefootmark{(a)} \\
Luminosity              & $L_*$ [$L_\odot$]          & 4.0\tablefootmark{(a)} \\
Effective Temperature   & $T_{\rm eff}$ [K]          & 4250\tablefootmark{(a)} \\
\hline
Disk & & \\
Disk mass               & $M_{\rm disk}$ [$M_\odot$] & 0.029\tablefootmark{(a,d)} \\
Surface density index   & $\gamma$                   & 0.9\tablefootmark{(a)} \\
Characteristic radius   & $R_{\rm c}$ [au]           & 46\tablefootmark{(a)} \\
Inner radius            & $R_{\rm in}$ [au]          & 0.14\tablefootmark{(a)} \\
Scale height index      & $\psi$                     & 0.11\tablefootmark{(a)} \\
Scale height angle      & $h_{\rm c}$ [rad]          & 0.18\tablefootmark{(a)} \\ 
\hline
Dust properties\tablefootmark{(e)} & & \  \\
Size                    & $a$ [$\mu$m]               & $0.005 - 1000$\tablefootmark{(a)} \\ 
Size distribution       &                            & $dn/da \propto a^{-3.5}$\tablefootmark{(a)} \\
Composition             &                            & ISM\tablefootmark{(a,b)} \\ 
\hline
Distance                & d [pc]                     & 125\tablefootmark{(c)} \\
Inclination             & $i$ [${}^\circ$]           & 20\tablefootmark{(c)} \\    
\hline
\end{tabular}
\tablefoot{\tablefoottext{a}{\cite{Andrews09}},\tablefoottext{b}{\cite{Draine84,Weingartner01}},\tablefoottext{c}{\cite{Pontoppidan11}},\tablefoottext{d}{Assuming a gas-to-dust ratio of 100},\tablefoottext{e}{See Online Figure \ref{fig:plot_opac}}}
\end{table}

The abundance structure of HCN is either constant or a jump (Figure \ref{fig:plot_disk_cbf}a) with a higher abundance in the hot FUV shielded inner disk ($x_{\rm in}$) and the outer disk ($x_{\rm out}$). In dense PDR models (e.g., \citealt{Meijerink05a}), HCN is abundant at ${\rm A}_{\rm V} >2$ and temperatures above $T_{\rm jump}=200$ K. We thus define the inner region by these two conditions, and later study the effect of changing the temperature threshold. Excitation effects are first studied with constant abundance models ($x_{\rm in}=x_{\rm out}$), before models with a jump in abundance are presented. The models include both thermal and turbulent broadening (FHWM $\sim 0.1$ km s$^{-1}$). Thermal broadening dominates in regions where the 3 $\mu$m and 14 $\mu$m emission emerges from.

The gas-to-dust (``G/D'') ratio is increased from the standard ratio of G/D=100 in some of the models, while the dust structure is kept the same for all models. As discussed in \cite{Meijerink09}, a higher gas-to-dust ratio leads to less dust absorption for high density gas. Thus, the line photosphere ($\tau_{\rm line}=1$) is shifted vertically upwards with respect to the dust photosphere ($\tau_{\rm dust}=1$) for an increased gas-to-dust ratio resulting in a higher line-to-continuum ratio. While we change the gas-to-dust ratio globally for the model, it does not mean that the disk gas mass changes by the same factor because an increased gas-to-dust ratio can be taken as a proxy for dust settling (\cite{Meijerink09}) and the IR lines trace only the inner part of the disk, while the bulk of the disk mass is in the outer disk.

Initially, large grains (0.005 $\mu$m - 1 mm) including grain growth are used, as adopted in the SED/image fitting by \cite{Andrews09}. In Section \ref{sec:disk_dustchange} the effect of smaller, ISM like, grains is studied. Similar to the gas-to-dust ratio, changing the dust opacity also varies the position of the dust photosphere ($\tau_{\rm dust}=1$) and the two effects are thus expected to be somewhat degenerate.

\subsection{The excitation mechanism of HCN} \label{sec:hcn_diskexcit}

The relative importance of excitation processes and formation of the R(6) $10^00-00^00$ 3 $\mu$m and Q(6) $01^10-00^00$ 14 $\mu$m lines is discussed with a model having a gas-to-dust ratio of 1000 and $x_{\rm in}=x_{\rm out}=3 \times 10^{-8}$. This model fits both the 3 $\mu$m CRIRES and the 14 $\mu$m \emph{Spitzer} observations reasonably well (Table \ref{tab:flx_constjump}).

\begin{figure*}[htb!]
\center
\sidecaption
\includegraphics[width=1.0\hsize]{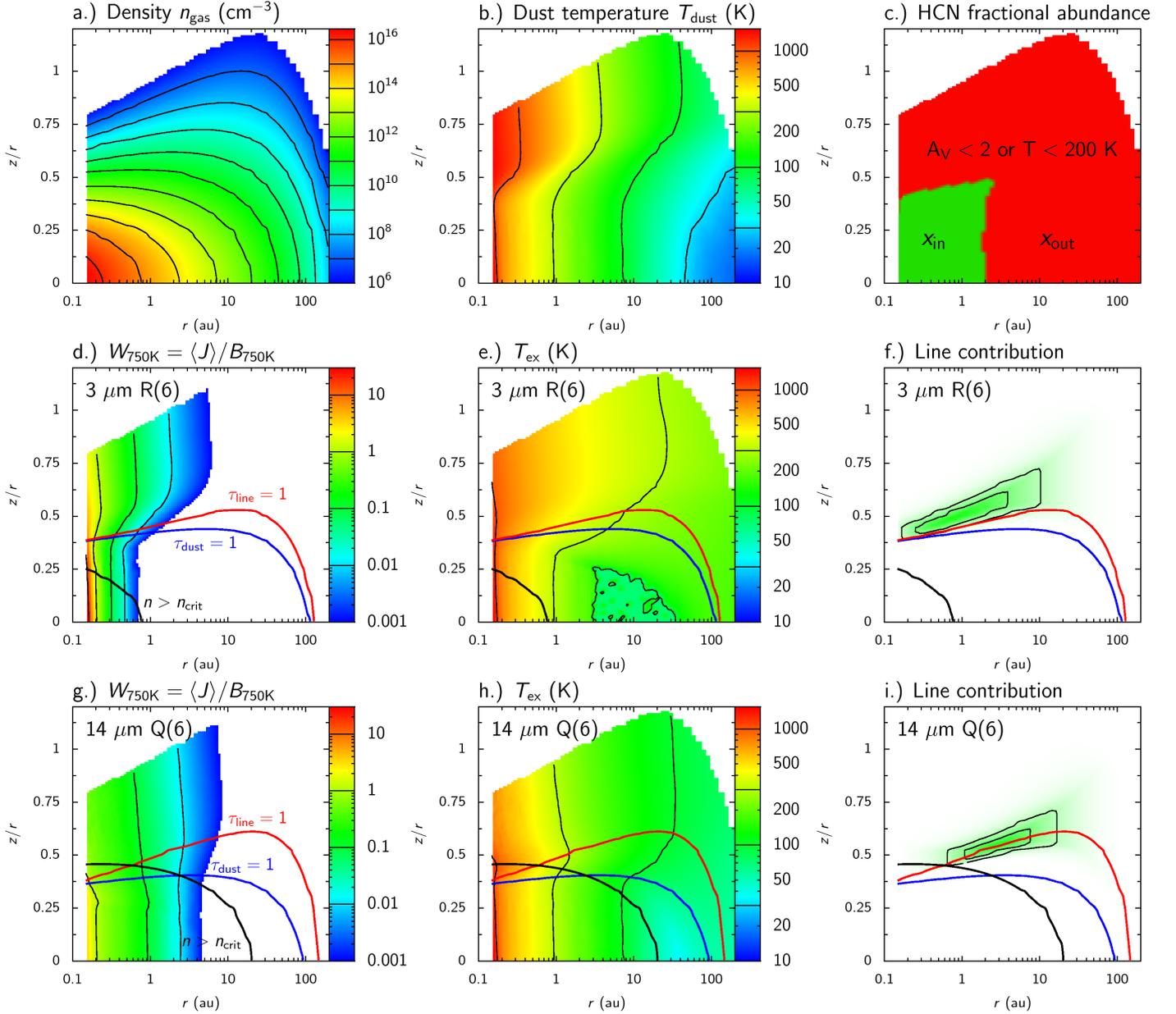}
\caption{Parameters of the AS 205 (N) model compared to the formation of the R(6) $10^00-00^00$ line at 3 $\mu$m and Q(6) $01^10-0^000$ line at 14 $\mu$m. The model with a gas-to-dust ratio of 1000 and a constant abundance $x_{\rm in}=x_{\rm out}=3 \times 10^{-8}$ is shown. \textbf{a)} Density structure. \textbf{b)} Dust temperature. \textbf{c)} Adopted fractional abundance of HCN. \textbf{d)} Continuum radiation field relative to a 750 K blackbody for the R(6) 3 $\mu$m line. Overlayed are the $\tau=1$ surface of the line (red) and continuum (blue) and the region where the critical density of the line is reached (black). \textbf{e)} Excitation temperature of the R(6) 3 $\mu$m line. \textbf{f)} Regions with 50 or 75 \% contribution to the observed flux. \textbf{g)-i)} As panels d)-f), but for the Q(6) 14 $\mu$m line.}
\label{fig:plot_disk_cbf}
\end{figure*}

Figure \ref{fig:plot_disk_cbf}a shows the density structure and Figure \ref{fig:plot_disk_cbf}b the temperature structure of the model. Densities up to a few times $10^{16}$ cm$^{-3}$ are reached in the inner midplane. The temperature spreads from $\sim 10$ K in the outer midplane up to the dust sublimation temperature of $\sim 1500$ K in the inner disk. Thus, there are region in the disk with a sufficiently high temperature and density to excite the HCN vibrational band collisionally, but these may be buried below the dust photosphere.

The vertical optical depth of the continuum and the line are indicated in Figures \ref{fig:plot_disk_cbf}d-\ref{fig:plot_disk_cbf}i by blue and red lines. The blue lines show the depth where the continuum gets optically thick ($\tau_{\rm dust}=1$) and the red lines where the HCN line does so ($\tau_{\rm line}=1$ at the line center). The black line indicates the height where the critical density of the transition is reached ($n > n_{\rm crit}$). Both of the 3 $\mu$m and 14 $\mu$m lines become saturated above the $\tau_{\rm dust}=1$ region, with peak line optical depth of about a few (3 $\mu$m) and $\sim 30$ (14 $\mu$m). The region where the 3 $\mu$ line is collisionally excited is entirely below $\tau_{\rm dust}=1$. For the 14 $\mu$m line the critical density is reached higher in the disk and above the $\tau_{\rm dust}=1$ in the inner 2 au. Thus, collisional excitation does not affect the 3 $\mu$m line but may affect the 14 $\mu$m line. 

Radiative pumping is studied in Figures \ref{fig:plot_disk_cbf}d and \ref{fig:plot_disk_cbf}g by the dilution factor relative to a $T_{\rm rad}=750$ K blackbody. Compared to the results in Section \ref{sec:depparam} (Figure \ref{fig:plot_paramspec}) where it is found that $W \sim 0.1 - 0.3$ are necessary for a detection within a region of $R=1$ au, the dilution factors are found to be lower. However, emitting regions out to 10 au are a factor of 100 larger and $W \sim 0.001$ can produce a considerable line emission since it approximately scales with $W$ (Figure \ref{fig:plot_3ratio14}). In fact, most of the IR continuum photons are produced in the innermost part of the disk and the continuum intensity in the upper disk decreases roughly as $\sim r^{-2}$, while the emitting region increases as $\sim r^2$. We may thus expect to find radiatively pumped molecules out to considerable distances. The HCN lines are seen in emission because the pumping photons are produced in the innermost part, while the radiatively excited molecules are mostly farther out and not in front of the continuum. This thus corresponds to the situation of \emph{resonant scattering}. 

The excitation temperature\footnote{Defined by $n_i/n_j=g_i/g_j \exp\left(-\Delta E/k T_{\rm ex}\right)$, with the level population $n_{i,j}$, the statistical weights $g_{i,j}$ and the energy difference $\Delta E$.} $T_{\rm ex}$ of the two lines is shown in Figures \ref{fig:plot_disk_cbf}e and \ref{fig:plot_disk_cbf}h. As a result of the strong radiation field, $T_{\rm ex} > T_{\rm dust}$ for most regions of the disk. Close to the midplane, in the region with $n > n_{\rm crit}$, the lines are thermalized by collisions and $T_{\rm ex} \sim T_{\rm dust}$. Both lines are thus excited out to large radii and regions beyond the innermost 1 au may contribute considerably to the emission.

Regions with the largest contribution to the lines are given in Figures \ref{fig:plot_disk_cbf}f and \ref{fig:plot_disk_cbf}i. Indicated is the region which contributes 50 or 75 \% of the observable line. Thus, most of the 3 $\mu$m line emerges from radial distances between the inner edge of the disk at 0.14 au and $\sim 10$ au. The 14 $\mu$m line emerges predominantly from about $1-15$ au, since the emission within the innermost 1 au cannot compete with the larger emitting region farther out. In the vertical direction, the emission of both lines is concentrated close to the $\tau_{\rm line}=1$ surface. Thus, the 14 $\mu$m emission also emerges mostly from regions with densities below the critical density, and radiative pumping dominates as for the 3 $\mu$m line. 

We conclude that radiative pumping is the major excitation mechanism for both the 3 $\mu$m and 14 $\mu$m lines studied here. As the size of the emitting region and geometrical dilution of the pumping IR field approximately balance out, the lines can be excited out to considerable distances on the order of 10 au. This is much larger than previous estimates of the emitting region based on LTE models and/or other molecules like water ($\sim 2.1$ au for 14 $\mu$m by \citealt{Salyk11} and $<1$ au for 3 $\mu$m by \citealt{Mandell12}).

\subsection{LTE versus non-LTE} \label{sec:lte_non_lte}

\begin{figure}[htb!]
\center
\includegraphics[width=0.75\hsize]{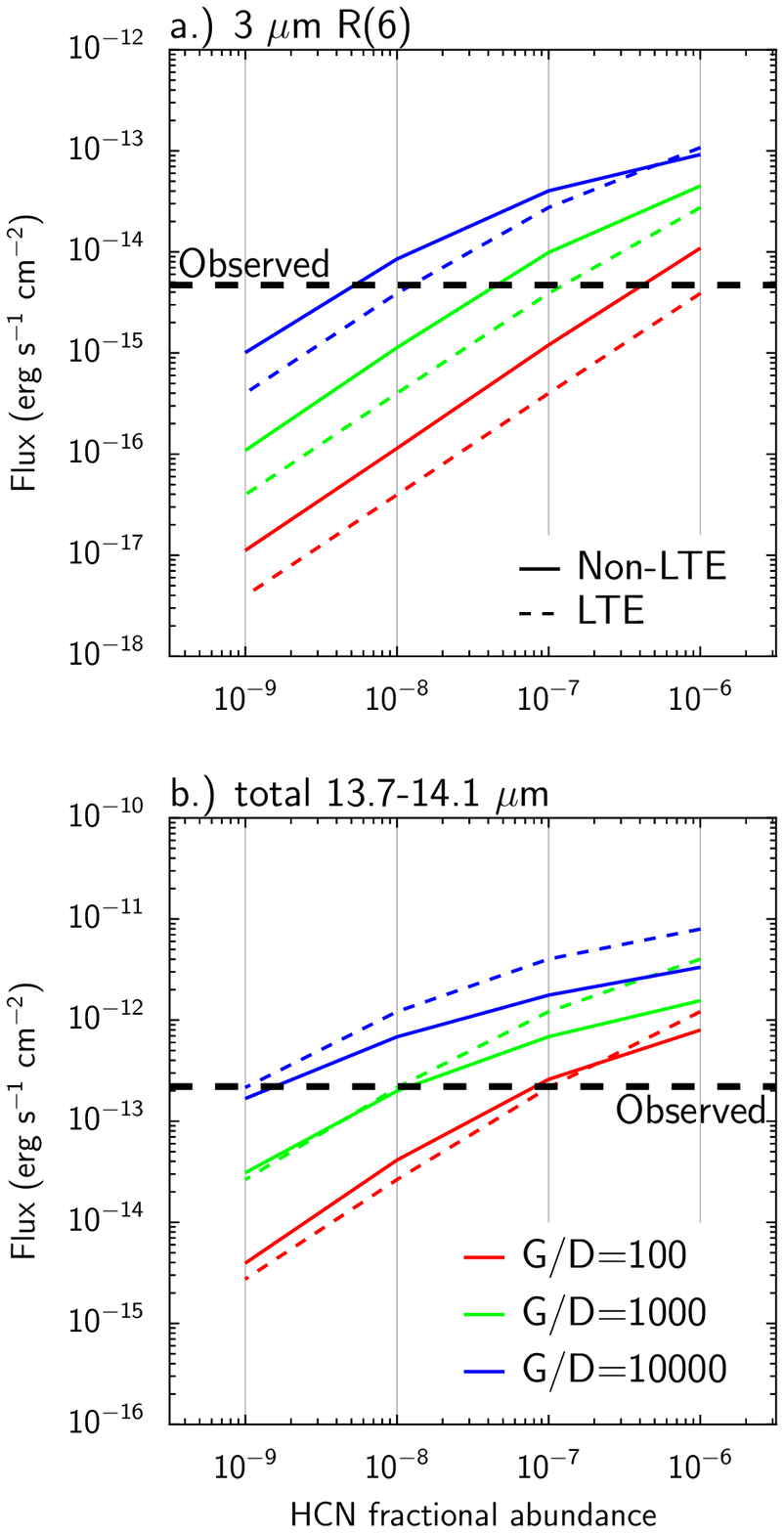}
\caption{Non-LTE and LTE results of the AS 205 (N) model for the \textbf{a)} R(6) $10^00-00^00$ line at 3 $\mu$m and \textbf{b)} the total HCN emission between 13.7 and 14.1 $\mu$m. The HCN fractional abundance is constant in the disk ($x_{\rm in}=x_{\rm out}$) and varied between $10^{-9}$ and $10^{-6}$. Three different gas-to-dust (G/D) ratios of 100 (red), 1000 (green), and 10000 (blue) are shown. Non-LTE results are shown in solid, LTE results in dashed line. The fluxes observed by \cite{Salyk09} (14 $\mu$m) and \cite{Mandell12} (3 $\mu$m) are given in black dashed lines.}
\label{fig:plot_diskflx_ltenlte}
\end{figure}

An important goal of this work is to find the bias in abundance determination introduced by the LTE assumption. In Figure \ref{fig:plot_diskflx_ltenlte}, the line flux of the R(6) $10^00-00^00$ line at 3 $\mu$m and the total emission between 13.7 and 14.1 $\mu$m of non-LTE models with varied HCN fractional abundance\footnote{Relative to the total gas density.} and gas-to-dust ratio are compared to LTE models. 

Somewhat surprisingly, the 3 $\mu$m line of the non-LTE model is stronger by a factor of up to $\sim 3$ compared to the LTE model. This is understood by the radiative pumping, which can excite the line superthermally ($T_{\rm ex} > T_{\rm dust}$) out to large radii of $\sim 10$ au. For the observed flux, the HCN abundance determined with a LTE model from the 3 $\mu$m line would be a factor of 3 too high. For the 14 $\mu$m total flux, the difference between non-LTE and LTE model are mostly smaller than for the 3 $\mu$m line ($<$ factor 2). As a result of the higher optical depth of the 14 $\mu$m line compared to the 3 $\mu$m line, the curve flattens off for higher abundance. At high HCN abundance, the lines cross and LTE models yield higher fluxes than the non-LTE model. This is understood by the $\tau_{\rm line}$ surface shifting to the upper disk for higher abundances. In that region, the 14 $\mu$m lines are slightly subthermally excited ($T_{\rm ex} < T_{\rm dust}$) as comparison of Figure \ref{fig:plot_disk_cbf}h with \ref{fig:plot_disk_cbf}b shows. The abundance determined from the 14 $\mu$m flux is almost unaffected by the LTE assumption.

\begin{figure*}[htb!]
\center
\includegraphics[width=1.0\hsize]{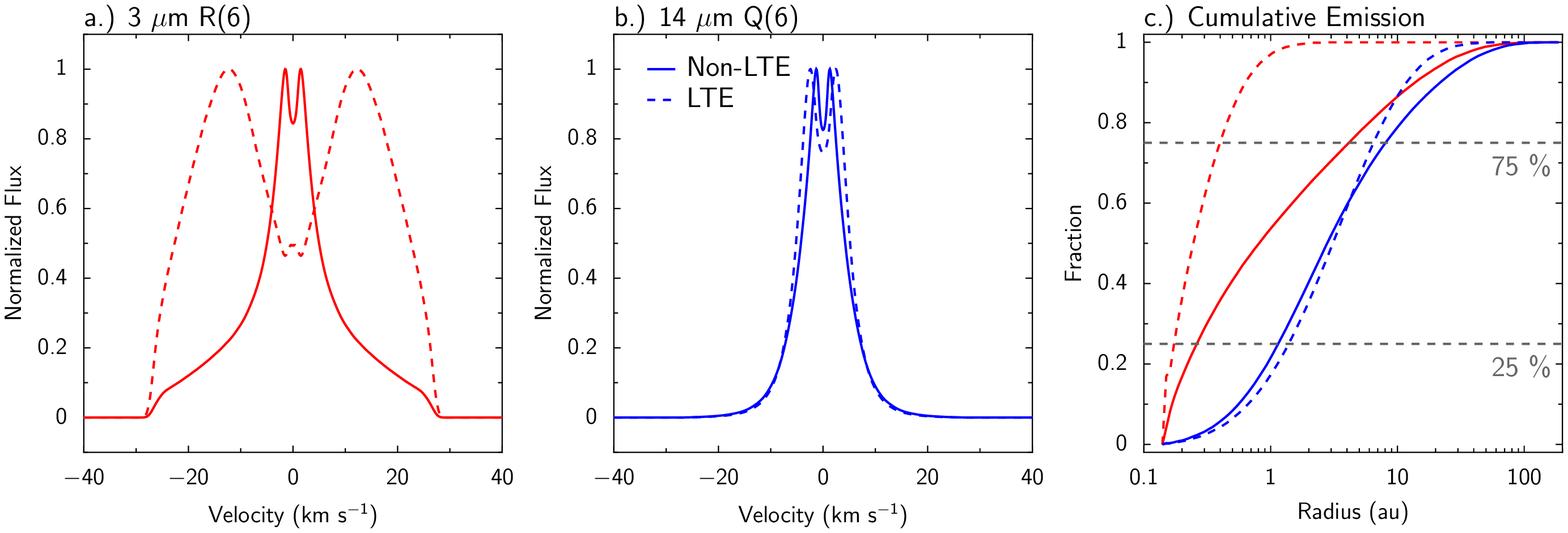}
\caption{Spectrum of the \textbf{a)} R(6) $10^00-00^00$ line at 3 $\mu$m and \textbf{b)} Q(6) $01^10-00^00$ line at 14 $\mu$m of the AS 205 (N) model with a gas-to-dust ratio of 1000 and a constant abundance of $x_{\rm in}=x_{\rm out}=3 \times 10^{-8}$. \textbf{c)} Contribution to the line flux within a certain radius. Results from the non-LTE and LTE model are shown in solid and dashed lines, respectively.}
\label{fig:plot_spec}
\end{figure*}

The gas-to-dust ratio and the HCN abundance are degenerate, which means that a factor of ten higher abundance but a factor of ten lower gas-to-dust ratio, resulting in the same total number of HCN molecules, yields almost the same flux (and line-to-continuum ratio). This would be different if collisional pumping is important for the excitation, because models with a high gas-to-dust ratio would then yield stronger lines due to the higher collisional partner density. Here, it means that the fractional abundance of HCN can only be determined relative to a gas-to-dust ratio. To detect lines, the line-to-continuum ratio can be the limiting factor (Appendix \ref{sec:app_lcratio}).

The line profiles of a LTE and non-LTE model with a gas-to-dust ratio of 1000 and $x_{\rm in}=x_{\rm out}=3 \times 10^{-8}$ are compared in Figure \ref{fig:plot_spec}. Also shown is the cumulative contribution to the flux within a certain radius. In the non-LTE model, both the 3 $\mu$m and 14 $\mu$m lines are narrow and peaky, with the 3 $\mu$m line having a broad component extending out to the maximum velocity of $\sim 27$ km s$^{-1}$. The peaky line profiles result from the large emitting regions of both lines, as the cumulative emission shows. The LTE models produce a much broader 3 $\mu$m line, while the 14 $\mu$m line is similar to the non-LTE model. The cumulative emission again reflects this. While similar regions contribute to the 14 $\mu$m emission in the LTE and non-LTE model, a much smaller region contributes in the LTE model to the 3 $\mu$m line. This is due to the larger upper level energy of the 3 $\mu$m line compared to the 14 $\mu$m line.
		
\subsection{Dependence on dust properties} \label{sec:disk_dustchange}

The degeneracy between the gas-to-dust ratio and molecular abundance, found in the previous section, shows that the total amount of visible HCN molecules determines the flux. Visible means that the molecules need to be above the dust photosphere. How does a change in grain opacity affect the lines? For comparison, we adopt the dust opacities by \cite{Mulders11} for a size distribution of 0.1-1.5 $\mu$m instead of the 0.005 $\mu$m - 1 mm large/grown grains used by \cite{Andrews09}. \cite{Mulders11} have used those opacities to fit the SED and images of HD 100546, a Herbig Be transitional disk. The smaller grains have an IR absorption similar the ISM grains, with an increased extinction opacity by a factor of 10 at 3 $\mu$m and 2.5 at 14 $\mu$m compared to the large grains (Online Figure \ref{fig:plot_opac}). While the models with changed dust opacity do not fit the AS 205 (N) SED anymore, they are illustrative to study the effects.

\begin{figure}[htb!]
\center
\includegraphics[width=0.75\hsize]{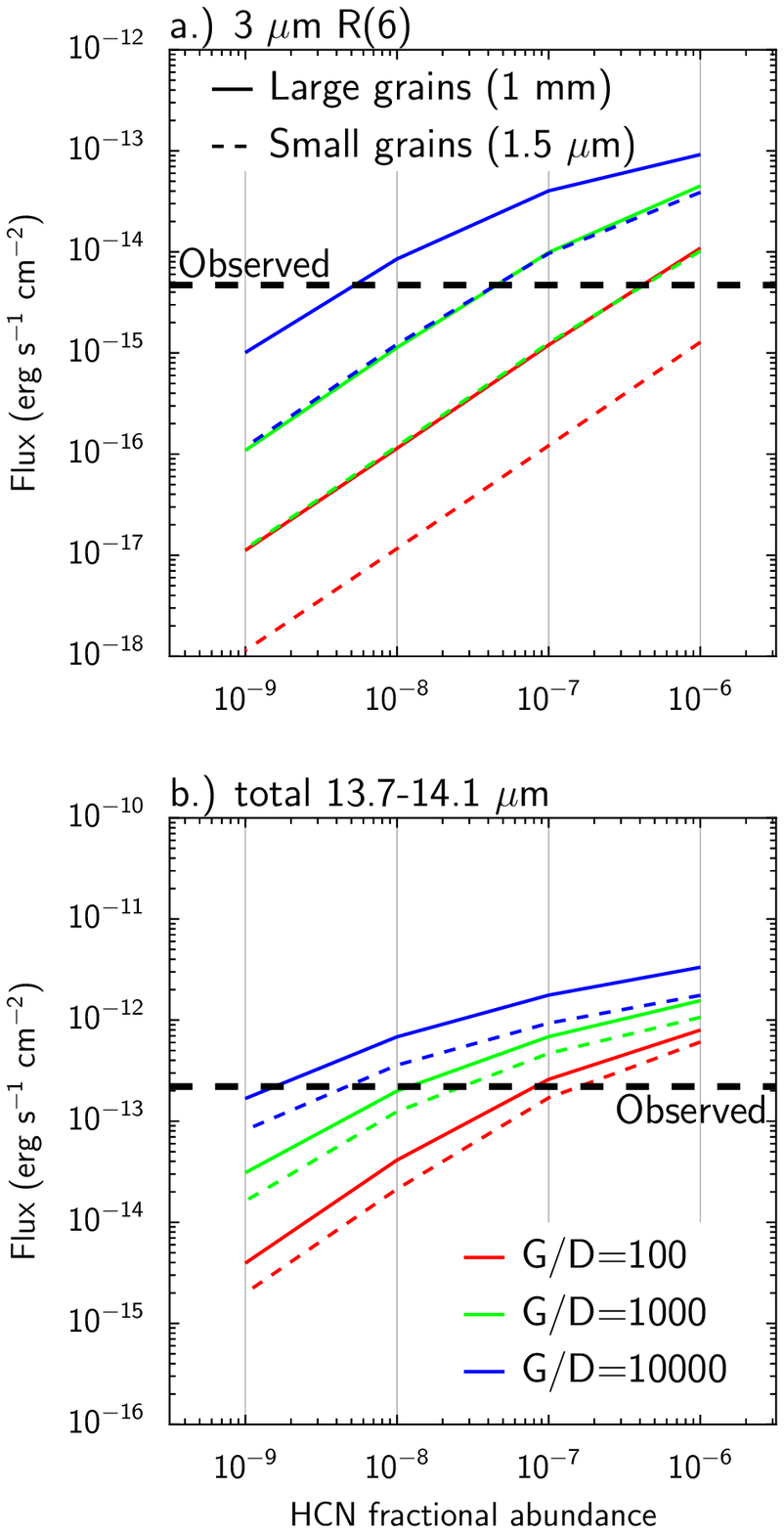}
\caption{Non-LTE results of the AS 205 (N) model comparing results with large grains (solid line) to ISM grains (dashed line). \textbf{a)} R(6) $10^00-00^00$ line at 3 $\mu$m and \textbf{b)} the total HCN emission between 13.7 and 14.1 $\mu$m. The adopted opacities are shown in Figure \ref{fig:plot_opac}. The HCN fractional abundance is constant in the disk ($x_{\rm in}=x_{\rm out}$) and varied between $10^{-9}$ and $10^{-6}$. Three different gas-to-dust (G/D) ratios of 100 (red), 1000 (green), and 10000 (blue) are shown. The fluxes observed by \cite{Salyk09} (14 $\mu$m) and \cite{Mandell12} (3 $\mu$m) are given in black dashed lines.}
\label{fig:plot_diskflx_dust}
\end{figure}

The fluxes of the 3 $\mu$m $10^00-00^00$ R(6) and total 13.7-14.1 $\mu$m emission for non-LTE models with both grain opacities are shown in Figure \ref{fig:plot_diskflx_dust}. The 3 $\mu$m lines scale almost exactly with the difference in the dust opacity. The 14 $\mu$m flux on the other hand, is scaled by a slightly smaller factor than the change in dust opacity, because of line opacity effects. Both the absolute fluxes and the 3 $\mu$m to 14 $\mu$m flux ratios are thus affected by the dust opacities. Thus, while changes in the dust opacities overall affect the line fluxes in a similar way as the gas-to-dust ratio, they may also change the ratios between vibrational bands. Here, the 14 $\mu$m / 3 $\mu$m flux ratios (Table \ref{tab:flx_constjump}) are increased by a factor of $\sim$3 to 6, away from the observed value.

\subsection{The effect of different abundance structures}

\begin{figure}[htb!]
\center
\includegraphics[width=0.75\hsize]{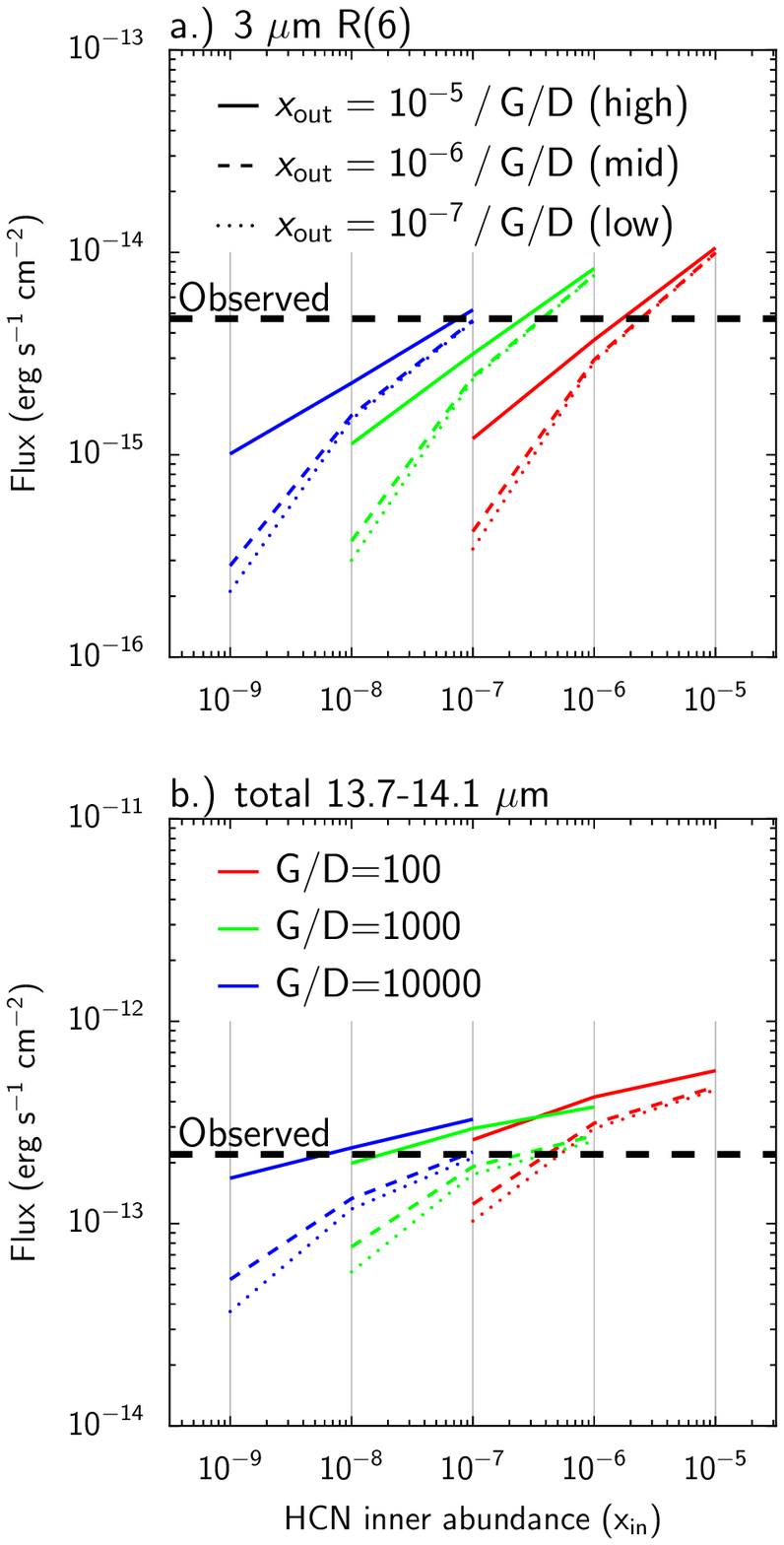}
\caption{Non-LTE results of the AS 205 (N) model comparing different abundances inside ($x_{\rm in}$) and outside ($x_{\rm out}$). \textbf{a)} R(6) $10^00-00^00$ line at 3 $\mu$m and \textbf{b)} the total HCN emission between 13.7 and 14.1 $\mu$m. The outer abundance is scaled for the gas-to-dust ratio (G/D) and takes 3 different values (e.g., $10^{-9}$, $10^{-8}$, and $10^{-7}$ for G/D$=100$). Three different gas-to-dust (G/D) ratios of 100 (red), 1000 (green), and 10000 (blue) are shown. The fluxes observed by \cite{Salyk09} (14 $\mu$m) and \cite{Mandell12} (3 $\mu$m) are given in black dashed lines.}
\label{fig:plot_diskflx_jump}
\end{figure}

Chemical models suggest that the HCN abundance is increased in the hot inner 1-2 au, due to a warm and hot-core like chemistry (\citealt{Lahuis00,Doty02,Agundez08,Bast13}). Non-LTE models with varying both inner ($x_{\rm in}$) and outer abundance ($x_{\rm out}$) are run to test the effects of such a ``jump abundance'' profile. In Figure \ref{fig:plot_diskflx_jump}, the 3 $\mu$m R(6) line flux and the total 14 $\mu$m emission is shown depending on the inner abundance and for different gas-to-dust ratios. The outer abundance is scaled such that the total number of HCN molecule in that region is the same for the three gas-to-dust ratios. Thus, the outer abundance is $x_{\rm out}=10^{-9..-7}$ for G/D$=100$ but $x_{\rm out}=10^{-11..-9}$ for G/D$=10000$.

The 3 $\mu$m lines depend more on the inner abundance than the 14 $\mu$m flux, due to the lower line optical depth and smaller emitting region. For the highest inner abundance of each gas-to-dust ratio, most of the emission at both 3 and 14 $\mu$m emerges from the inner region ($<2$ au), since the line fluxes do not change much with $x_{\rm out}$ anymore.

For both a gas-to-dust ratio of 10000 (with $x_{\rm in}=10^{-7}$ and $x_{\rm out}=10^{-10}$) and a gas-to-dust ratio of 1000 (with $x_{\rm in}=3 \times 10^{-7}$ and $x_{\rm out}=10^{-9}$), the observed fluxes can be reproduced well (Table \ref{tab:flx_constjump}). No combination of $x_{\rm in}$ and $x_{\rm out}$ can be found to equally well reproduce both fluxes when the gas-to-dust ratio is set to 100. This is because of the lower density for lower gas-to-dust ratios in the inner 2 au close to the $\tau_{\rm dust}=1$ surface. The densities in that region are above the critical density for higher gas-to-dust ratios as Figure \ref{fig:plot_disk_cbf} shows. Thus, in the situation where most of the 14 $\mu$m emission emerges from the inner most region due to the chemical abundance structure, collisional excitation can become important. Consequently, the degeneracy between gas-to-dust ratio and HCN abundance is broken.

\begin{table*}[tbh]
\caption{Fluxes of the R(6) $10^00-00^00$ line at 3 $\mu$m and the total HCN emission between 13.7 and 14.1 for AS 205 (N) models with a constant or jump abundance compared to observations.}
\label{tab:flx_constjump}
\centering
\begin{tabular}{lcccc|ccc}
\hline\hline
Abundance profile & $T_{\rm jump}$ & G/D ratio &  $x_{\rm in}$ & $x_{\rm out}$ & 3 $\mu$m & 14 $\mu$m & Flux ratio \\
&       [K]     &         &             &               &  \multicolumn{2}{c}{[erg s$^{-1}$ cm$^{-2}$]} & 14 $\mu$m / 3 $\mu$m \\\hline
Constant    & -      & 1000    & 3(-8) & 3(-8)  & 3.3(-15) & 3.8(-13) & 115 \\
Jump        & 200    & 1000    & 3(-7) & 1(-9)  & 4.7(-15) & 2.4(-13) & 51  \\
Jump        & 200    & 10000   & 1(-7) & 1(-10) & 4.6(-15) & 2.3(-13) & 50  \\
Jump        & 300    & 1000    & 6(-7) & 1(-10) & 6.0(-15) & 1.6(-13) & 27  \\
Jump        & 300    & 10000   & 6(-7) & 1(-11) & 4.4(-15) & 2.0(-13) & 45  \\
\hline
\multicolumn{2}{l}{Observed}   &     &       &        & 4.7(-15) & 2.2(-13) & 46\\
\hline
\end{tabular}
\tablefoot{$a(b)$ means $a \times 10^{b}$}
\end{table*}

\begin{figure*}[htb!]
\center
\includegraphics[width=1.0\hsize]{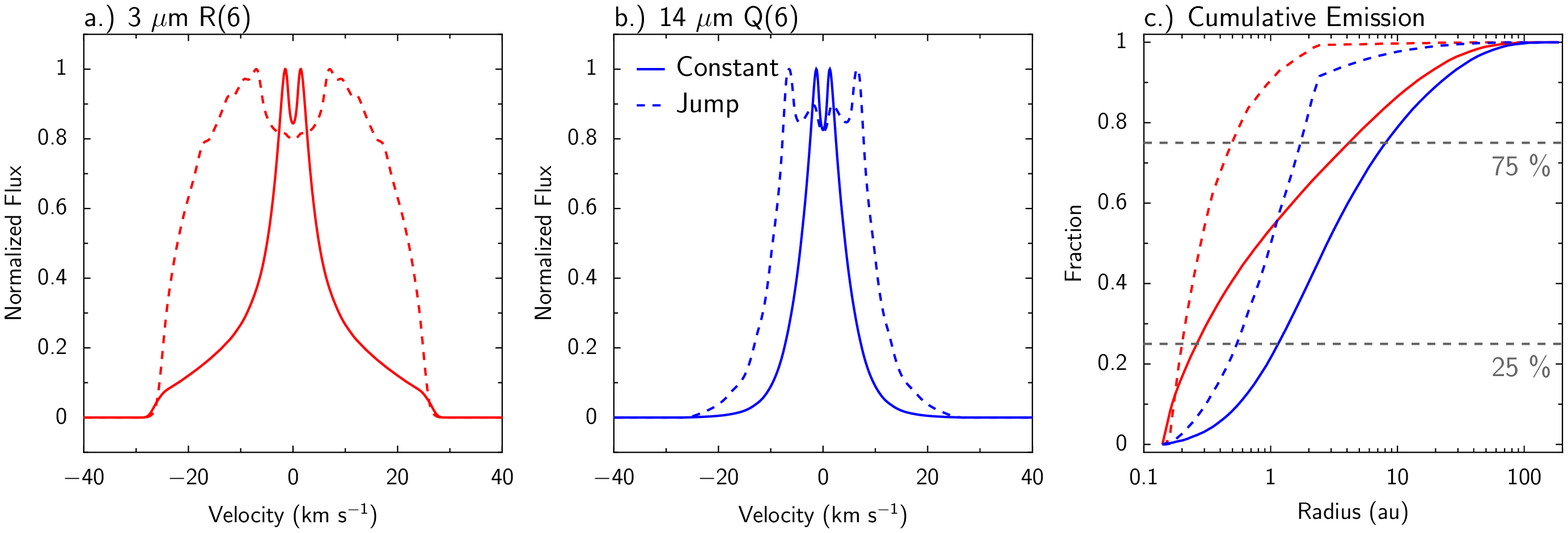}
\caption{Spectrum of the \textbf{a)} R(6) $10^00-00^00$ line at 3 $\mu$m and \textbf{b)} Q(6) $01^10-00^00$ line at 14 $\mu$m of the AS 205 (N) model with constant abundance (G/D$=1000$ and $x_{\rm in}=x_{\rm out}=3 \times 10^{-8}$, solid lines) or with jump abundance (G/D$=1000$, $x_{\rm in}=3 \times 10^{-7}$, and $x_{\rm out}=1 \times 10^{-9}$, dashed lines). \textbf{c)} Contribution to the line flux within a certain radius.}
\label{fig:plot_spec_jump}
\end{figure*}

Most of the emission in the well-fitting jump abundance models is from the innermost 2 au, because of the large jump in abundance. This is unlike the constant abundance models with emission out to $\sim 10$ au. The effect of the different emitting regions on the line profile is shown in Figure \ref{fig:plot_spec_jump}. As expected, both the 3 $\mu$m and 14 $\mu$m lines from the jump abundance models are much broader compared to the constant abundance models. The cumulative emission panel indeed confirms that the emission is much more centered in both lines and almost the entire emission of the 3 $\mu$m line and 75 \% of the 14 $\mu$m line emerges from inside 2 au. While the line shape of the 3 $\mu$m line with jump abundance (Figure \ref{fig:plot_spec_jump}) and LTE excitation (\ref{fig:plot_spec}) are somewhat similar, one is the result of the abundance structure and the other of the excitation.

\subsection{The size of the inner region} \label{sec:spitzer}

So far, we have assumed that the inner abundance extends out to regions with $T > T_{\rm jump}=200$ K. Increasing $T_{\rm jump}$ leads to a smaller, but warmer, emitting region. To illustrate the effect, we show models with $T_{\rm jump}=300$ K in Table \ref{tab:flx_constjump}. To still fit the observed line fluxes, the inner abundance needs to be increased, while the outer abundance lowered. Since the emitting region is warmer, the rotational temperature of the lines is higher, as it is mostly determined by collisional excitation of the vibrational ground state.

\begin{figure*}[htb!]
\center
\sidecaption
\includegraphics[width=0.75\hsize]{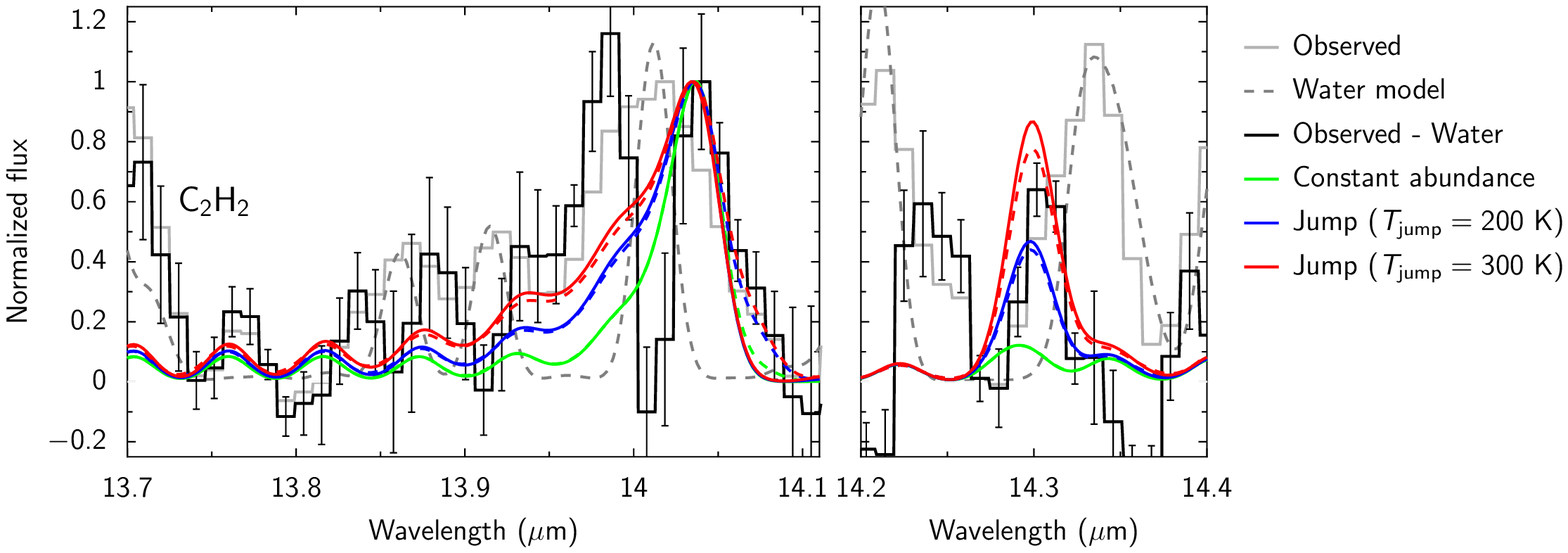}
\caption{\emph{Spitzer} observations compared to the models given in Table \ref{tab:flx_constjump} with G/D$=1000$ and constant or jump abundance. Dashed color lines show models including H$^{13}$CN. Gray lines show the observed spectrum and the best-fit water model by \cite{Salyk11}. The black line indicates the water subtracted spectrum. The spectra are normalized to the peak of the Q-branch at around 14.05 $\mu$m.}
\label{fig:plot_spitzer}
\end{figure*}

Figure \ref{fig:plot_spitzer} compares the \emph{Spitzer} HCN features of the Q-branches of the $01^10-00^00$ (14.05 $\mu$m) and $02^00-01^10$ (14.3 $\mu$m) bands to three models convolved to the \emph{Spitzer} spectral resolving power of $R \sim 600$. Data by the c2d legacy program is used (\citealt{Lahuis06,Evans09}). Models with G/D$=1000$ and constant abundance or jump abundance with varied $T_{\rm jump}$ are shown. Also shown are models with a contribution of H$^{13}$CN included. These models are calculated assuming $60=[{}^{12}{\rm C}] / [{}^{13}{\rm C}]$. Unfortunately, both HCN features overlap with water lines. Thus, the best fit water model by \cite{Salyk11} is subtracted before comparison. 

We find that the Q-branch at 14.05 $\mu$m is best reproduced by the model with $T_{\rm jump}=300$ K, however some residuals at around 14.0 $\mu$m remain, possibly because of an overcorrection of the water emission. The $02^00-01^10$ Q-branch at 14.3 $\mu$m is stronger in the models with jump abundance and overproduces observations for $T_{\rm jump}=300$ K. While a model with jump abundance seems to fit observations better, none of the models can well reproduce both Q-branches. The comparison is however hampered by uncertainties in the correction for water emission. Interestingly, the red wing of the 14.05 $\mu$m Q-branch is better reproduced in the models with H$^{13}$CN included.

%
%

\section{Discussion} \label{sec:discussion}

\subsection{Excitation and line formation effects}

Owing to the very high critical densities of the HCN ro-vibrational lines, it could be expected that the lines are subthermally excited, resulting in large differences in the abundance determined by LTE or non-LTE models. However, since infrared pumping provides an additional way of exciting the molecules, we find that the fluxes between LTE and non-LTE models do not differ by more than a factor of few and consequently the abundance determination using LTE or non-LTE models is only affected by about the same factor. However, the line profiles between LTE and non-LTE models can differ considerably, since IR pumping can efficiently pump molecules out to several au. This indicates that if very small emitting regions are found from observations, they may be attributed to changes in the chemistry rather than excitation effects.

Infrared pumping results in a narrow and ``peaky'' line profile, e.g. of the 3 $\mu$m $10^00-0000$ R(6) line, which are somewhat similar to the line profiles found by \cite{Bast11} in velocity resolved observations of CO ro-vibrational lines. The CO line profiles do not have a central dip as HCN, maybe due to line opacity effects (Figure 7 in \citealt{Elitzur12}) or the spectral resolution. For CO, spectro-astrometry constrains the emitting region to be smaller than a few au and a disk wind has thus been employed to explain the line profile (\citealt{Pontoppidan11}). Since the HCN lines observed by CRIRES are much weaker than the CO lines, spectro-astrometry cannot be used for HCN and the line profiles are too noisy for a direct comparison. Since it is unclear, if HCN follows CO, we cannot currently distinguish between a disk wind scenario and extended emission due to IR pumping for HCN. Our HCN results however show that IR pumping can also result in peaky profiles. 

Both models with constant HCN abundance or jump abundance with an elevated inner and lowered outer abundance can simultaneously reproduce the 3 $\mu$m and 14 $\mu$m observations reasonably well. The jump abundance models however yield much broader line profiles both at 3 $\mu$m and 14 $\mu$m. Comparing non-LTE with LTE models, only an increase in line width of the 3 $\mu$m line was found. Thus, to observationally distinguish between excitation and abundance structure effects based only on line profiles, both emission bands need to be velocity resolved. Constraining the position of the abundance jump ($T_{\rm jump}$) will help to test if the size of the emitting region of water and HCN is similar. Different assumptions about the size of the HCN region can lead to more than an order of magnitude different column densities and thus HCN/H$_2$O abundance ratios (e.g., \citealt{Salyk11}, \citealt{Carr11}). If the size of the HCN emitting region proves to be similar to that of H$_2$O for sources where the line profiles can be spectrally resolved, this would greatly facilitate future analyses of large samples of sources with spectrally unresolved data, such as observed with Spitzer and expected with MIRI.

Since collisional excitation is not important for most models, the gas-to-dust ratio and HCN fractional abundance are degenerate and only the absolute amount and not the fractional abundances of HCN can be determined. While this degeneracy does not alter the ratio between the 3 $\mu$m and 14 $\mu$m lines, changes in the dust opacity (e.g., through grain growth) can also change the line ratio. The degeneracy between gas-to-dust ratio and HCN fractional abundance is broken for models where most of the emission is from the innermost 2 au because of the adopted abundance profile. Then, collisional excitation can be important for the 14 $\mu$m line. For AS 205 (N), this yields a constraint for the gas-to-dust ratio $>100$. This constraint is thus explained by a different effect than that found by \cite{Meijerink09} for water. For water, the fractional abundance is fixed as all oxygen is driven into water in hot gas ($T > 250$ K), and the gas-to-dust ratio needs to be increased until a large enough water column density is above the $\tau_{\rm dust}=1$ surface to yield the observed line-to-continuum ratio. In contrast to water, the HCN abundance is not fixed by an elemental abundance and is a \emph{free parameter}. Thus, the observed line-to-continuum ratio of HCN can be reproduced by adjusting the fractional abundance and does not drive to the conclusion of a high gas-to-dust ratio as in water found by \cite{Meijerink09}.

The ro-vibrational lines of HCN show considerably different excitation effects than water (\citealt{Meijerink09}). Radiative pumping is far more important for HCN, since the upper levels of the observable lines can be directly radiatively pumped from the vibrational ground-state in which most molecules are throughout the disk. In water, however, mostly pure rotational lines are observed and selection rules force radiative pumping through steps up the ladder and collisional pumping will be more important. Thus, the excitation of the vibrational lines of acetylene (C$_2$H$_2$) near 14 $\mu$m is likely similar to HCN. Indeed, non-LTE models of C$_2$H$_2$ by \cite{Lacy13} show the importance of radiative pumping. As for HCN, he finds that the level population is not too far from the LTE. It is thus unlikely that the trends found between HCN and C$_2$H$_2$ (\citealt{Pascucci09}) can be entirely attributed to excitation effects. Following this comparison between HCN and C$_2$H$_2$, the level structure of CO$_2$ (e.g., \citealt{GonzalezAlfonso99}, Figure 5) suggests that also its excitation is predominantly by radiation. It should be noted, though, that this comparison of HCN with C$_2$H$_2$ and CO$_2$ is only valid if collisional excitation rates are similar, since collisional excitation has some importance for the 14 $\mu$m lines of HCN in jump abundance models. The CO$_2$-H$_2$ rate coefficient for the collisional de-excitation of the $\nu_2$ bending mode is on the order of $6 \times 10^{-12}$ cm$^3$ s$^{-1}$ (see \citealt{Boonman03c} and \citealt{Allen80}) similar to the $\nu_2$ bending mode of HCN-H$_2$ ($1.3 \times 10^{-11}$ cm$^3$ s$^{-1}$). Also the C$_2$H$_2$-H$_2$ rate coefficient for the $\nu_5$ bending mode is with $3 \times 10^{-12}$ cm$^3$ s$^{-1}$ (\citealt{Hager81}) on the same order of magnitude.

\subsection{Abundances and comparison to chemical models}

The line fluxes are not affected in a major way by the LTE assumption and our derived fractional abundances do not substantially differ from previous estimates. For AS 205 (N), \cite{Mandell12} derive a HCN fractional abundance of $2 \times 10^{-7}$ relative to H$_2$ from only the 3 $\mu$m lines. They used an LTE model with a gas-to-dust ratio of 12800 and smaller grains, similar to those discussed in Section \ref{sec:disk_dustchange}. Our LTE model for the same abundance, gas-to-dust ratio of 12850 and small grains, yields a flux of $3.9 \times 10^{-15}$ erg s$^{-1}$ cm$^{-2}$ for the 3 $\mu$m R(6) line, within about 20 \% to the observed flux, confirming the agreement between our and their approach.

The derived fractional abundances (Table \ref{tab:flx_constjump}) are within the range predicted by chemical models (e.g., \citealt{Markwick02,Agundez08,Willacy09,Walsh10,Najita11,Walsh12,Walsh14}). As noted by \cite{Mandell12}, they predict a HCN to H$_2$O fractional abundances in the warm ($T \gtrsim 200$ K) gas of $10^{-4} - 10^{-1}$, leading to HCN fractional abundances between $10^{-8}$ and $10^{-5}$. In colder regions, the HCN abundance is predicted to be much lower (e.g., Figure 1 in \citealt{Agundez08}) and our models with jump abundance profiles indeed require $x_{\rm in} > x_{\rm out}$. While for AS 205 (N), it is difficult to distinguish between constant and jump abundance profiles with the current spectrally unresolved data (Section \ref{sec:spitzer}), it is interesting that other disks point to a jump abundance (\citealt{Pontoppidan14}).

\subsection{Predictions for JWST-MIRI, METIS, and ALMA}

\begin{figure*}[htb!]
\center
\includegraphics[width=0.8\hsize]{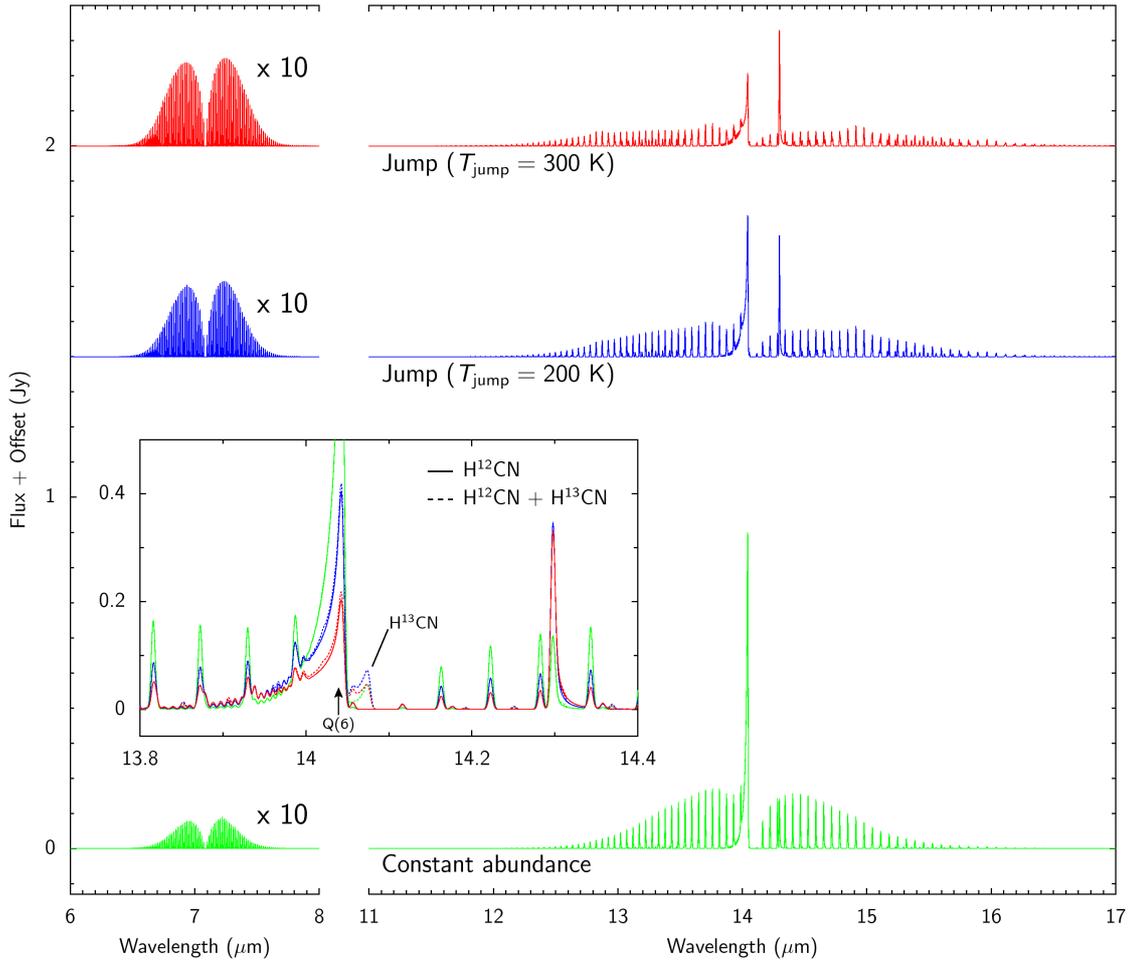}
\caption{Predictions for JWST-MIRI. The models given in Table \ref{tab:flx_constjump} with G/D$=1000$ and constant or jump abundance are shown, convolved to the spectral resolving power of $R=3000$. The inset also shows a spectrum of both H$^{12}$CN and H$^{13}$CN (dashed line).}
\label{fig:plot_miri}
\end{figure*}

The MIRI instrument on JWST (\citealt{Wright04}) will have a substantially better sensitivity and higher spectral resolution ($R=3000$) compared to \emph{Spitzer} ($R=600$). This will allow us to detect resolved single lines of the 7 $\mu$m and 14 $\mu$m P- and R-branches. The Q-branches will however still be seen as a blend of several lines. Figure \ref{fig:plot_miri} shows MIRI predictions of the three models with G/D$=1000$ and constant or jump abundance. Abundances given in Table \ref{tab:flx_constjump} are used. The inset in the figure magnifies the Q-branches at 14 $\mu$m and also includes H$^{13}$CN, calculated under the assumption of an abundance ratio $60=[{}^{12}{\rm C}] / [{}^{13}{\rm C}]$. The line detection limit of MIRI\footnote{http://www.stsci.edu/jwst/instruments/miri/sensitivity/, \mbox{$\sim 10^{-20}$ W m$^{-2}$ after $10^4$ s}} corresponds to line peaks of a few mJy for $R=3000$ even after a minimum exposure time of only 30 seconds and all lines that can be recognized in the figure are easily detected, provided that a high enough signal-to-noise can be reached on the continuum (Appendix \ref{sec:app_lcratio}).

MIRI predictions of the three models differ in a similar way as found with the \emph{Spitzer} spectra in Section \ref{sec:spitzer}. Since the jump abundance models emit from warmer regions, the P- and R-branches at 14 $\mu$m emit to higher-$J$ transitions compared to the model with constant abundance. The $02^00$-levels are more populated close to the star and thus both the Q-branch at 14.3 $\mu$m and the 7 $\mu$m lines are stronger in the jump abundance models. Differences between the models can be seen better in rotational diagrams (Online Figure \ref{fig:plot_rot}). The constant abundance model exhibits a curved shape over the whole range of transitions because of different radii probed. The jump abundance models have higher abundances in the inner region and the lines thus become optically thick, seen in a strongly curved shape of the lower-$J$ lines. The H$^{13}$CN isotopologue is predicted to be easily detected and identified by its Q-branch at $\sim 14.05$ $\mu$m, which is sufficiently shifted in wavelength not to be blended with the main isotopologue. We conclude that thanks to high sensitivity and spectral resolution of MIRI, it will be possible to distinguish between models with different abundance structures and to directly study the chemistry in regions were terrestrial planets form. 

The METIS instrument on the E-ELT (\citealt{Brandl14}) will be able to observe at 3-14 $\mu$m with a spectral resolving power of $R=10^5$ at 3 $\mu$m and $R=5000$ at 14 $\mu$m. In the Online Figure \ref{fig:plot_metis}, the line profiles of the 3 $\mu$m R(6) and 14 $\mu$m Q(6) lines from the same models as in Figure \ref{fig:plot_miri} are shown. At 14 $\mu$m, the velocity profile cannot be resolved, but single lines of the P- and R-branches can be seen, similar to MIRI. The factor of 100 better sensitivity of METIS compared to CRIRES will allow us to easily detect the 3 $\mu$m lines of HCN velocity resolved. These velocity resolved observations with a good signal-to-noise ratio will simplify to constrain the position of the abundance jump ($T_{\rm jump}$) and to better quantify the inner disk chemistry.

ALMA cannot resolve the inner few au, but several submillimeter lines within vibrationally excited states emerge from that region and are in the wavelength range of ALMA receivers. Compared to envelopes of protostars, where these submillimeter lines are regularly detected, the size of the emitting regions is very small in disks. We find that the most promising line for a detection is the $J=4-3$ line within the $01^10$ level (356.256 GHz). For the three models with G/D$=1000$ and constant or jump abundance (Table \ref{tab:flx_constjump}), the total line fluxes vary between 82 mJy km s$^{-1}$ (constant), 26 mJy km s$^{-1}$ ($T_{\rm jump}=200$ K) and 10 mJy km s$^{-1}$ ($T_{\rm jump}=300$ K). For a one hour integration with 50 12m antennas and 2nd quantile weather, the $1\sigma$-sensitivity of ALMA is 6 mJy km s$^{-1}$. Thus, while the constant abundance model can be detected within one hour, the jump abundance models require a longer integration time for a detection. The Online Figure \ref{fig:plot_alma} shows the spectra of the three models together with the detection limit for 5 km s$^{-1}$ channels within 3 hours indicated. Depending on the abundance structure, ALMA may thus be able to detect lines between vibrationally excited levels and thus probe the inner few au of a protoplanetary disk.

%
%

\section{Conclusions} \label{sec:conclusions}

We have modeled the non-LTE excitation of ro-vibrational lines of the organic molecule HCN, a chemically important ingredient of protoplanetary disks. Single point slab models to study the relative importance of different excitation effects were calculated and the emission of the T Tauri disk AS 205 (N) was modeled. That disk was observed in different vibrational bands by VLT-CRIRES at 3 $\mu$m and \emph{Spitzer} at 14 $\mu$m. Models with different abundance structures, gas-to-dust ratios and dust opacities were run to explore the effect on the ro-vibrational lines. The main conclusions of this study are:

\begin{itemize}
\item The ro-vibrational lines of HCN have very high critical densities and collisional excitation alone can only reproduce the observed ratio between 3 and 14 $\mu$m lines for density $> 10^{13}$ cm$^{-3}$ and temperatures $> 750$ K. Radiative pumping by IR photons can easily reproduce the observed line ratio.
\item Infrared pumping can excite both the 3 and 14 $\mu$m lines out to considerable radii of $\sim 10$ au. It may thus point to a chemical rather than an excitation effect, if smaller emitting regions are found (e.g., by spatially or velocity resolved observations).
\item Fluxes derived from LTE or non-LTE models differ not more than about a factor of $\sim 3$ as a result of the IR pumping of HCN. Consequently, the abundances derived by LTE and non-LTE models do not differ by a larger factor. The line profiles of the 3 $\mu$m lines can however differ considerably.
\item Both models with a constant abundance and a jump abundance, as expected form chemical considerations, can reproduce the observed fluxes reasonably well. The line profiles and extent of the emitting area however differ considerably. While it is difficult to distinguish between the scenarios with current data, it will be easily possible with JWST-MIRI and E-ELT METIS observations. Depending on the scenario, ALMA can detect rotational lines within vibrationally excited levels.
\item The gas-to-dust ratio and fractional abundance of HCN are degenerate, since collisional excitation is mostly unimportant. The adopted dust opacity may change the line fluxes in a similar way but can also alter the ratio between the 3 and 14 $\mu$m lines.
\end{itemize}

Our work shows that despite the very high critical densities of ro-vibrational lines, the LTE assumption does not necessarily lead to large difference in the derived abundances, because of radiative pumping by the continuum radiation field. This will facilitate future studies of warm inner disk chemistry considerably, since other key molecules (C$_2$H$_2$ and CO$_2$) are also expected to be excited in a similar way.

\begin{acknowledgements}
We thank Jeanette Bast, Alan Heays, Fred Lahuis, Klaus Pontoppidan, Ian Smith, Xander Tielens, and an anonymous referee for useful discussions and Colette Salyk for sharing her water model of AS 205 (N). Astrochemistry in Leiden is supported by the Netherlands Research School for Astronomy (NOVA), by a Royal Netherlands Academy of Arts and Sciences (KNAW) professor prize, and by the European Union A-ERC grant 291141 CHEMPLAN.
\end{acknowledgements}

\bibliographystyle{aa}

\Online 

\begin{appendix}

\section{Online figures}

\begin{figure}[htb!]
\center
\includegraphics[width=1.0\hsize]{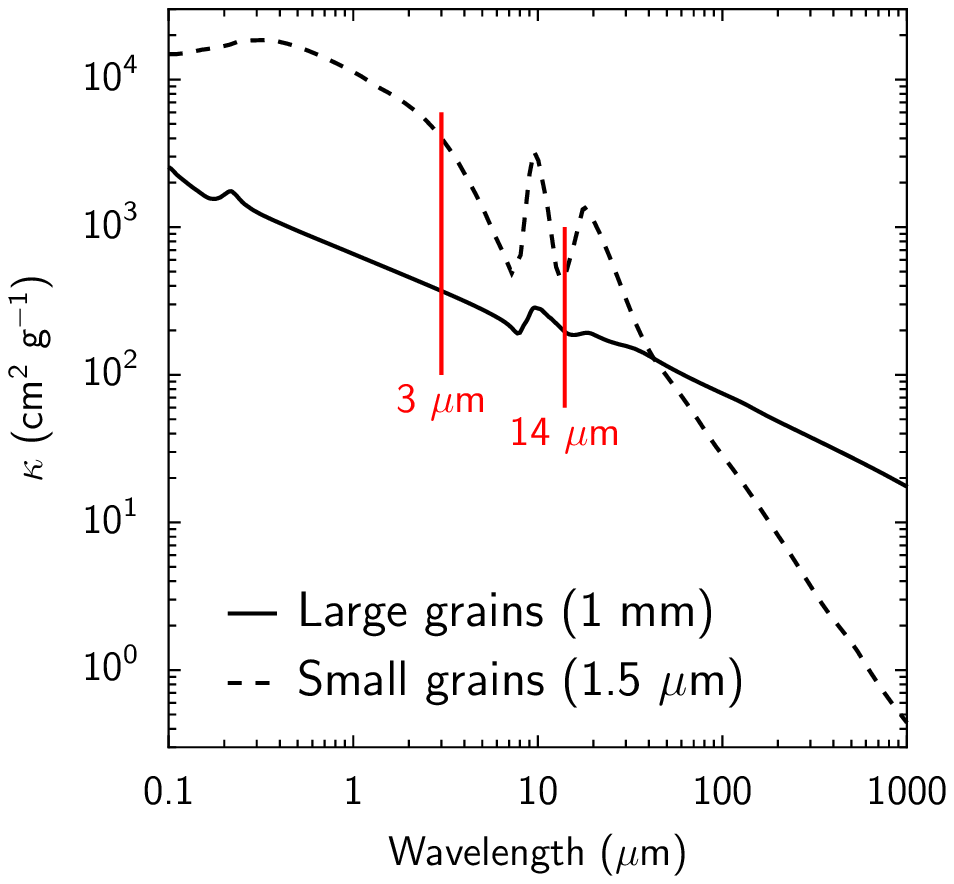}
\caption{Adopted mass opacities for large vs. small ISM grains. The extinction opacity (absorption$+$scattering) is shown.}
\label{fig:plot_opac}
\end{figure}

\begin{figure*}[htb!]
\center
\includegraphics[width=1.0\hsize]{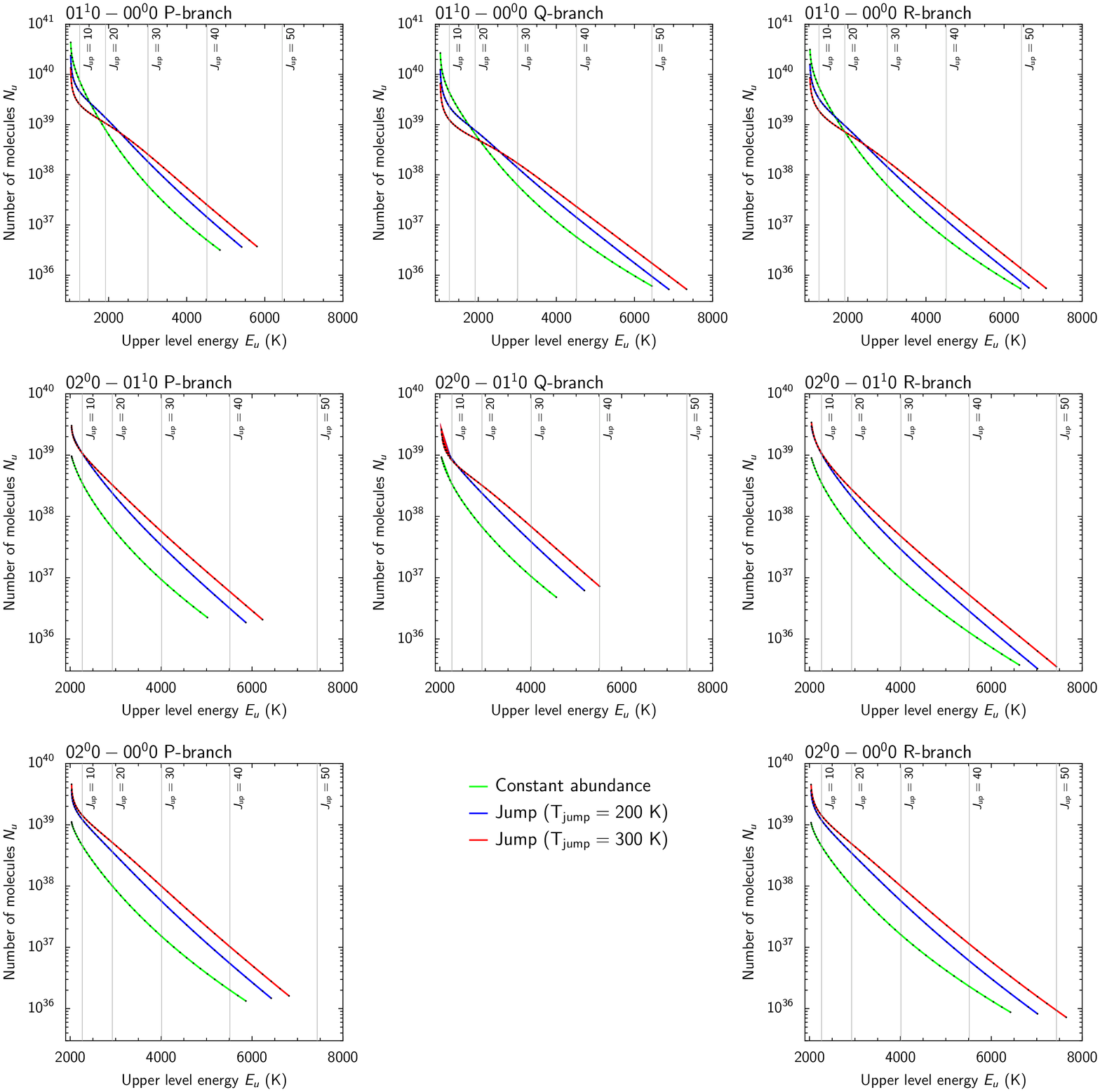}
\caption{Rotational diagrams of the models given in Table \ref{tab:flx_constjump} with G/D$=1000$ and constant or jump abundance. Only lines above the detection limit of MIRI within 10000 seconds ($\sim 10^{-20}$ W m$^{-2}$) are shown. The $x$-axis shows the upper level energy $E_u$ (K) and the $y$-axis the number of molecules, $N_{\rm u}= 4 \pi d^2 F / (A_{ul} h \nu_{ul} g_u)$, where $d$ is the distance, $F$ flux, $A_{ul}$ Einstein-$A$ coefficient, $\nu_{ul}$ transition frequency, and $g_u$ the upper level degeneracy.}
\label{fig:plot_rot}
\end{figure*}

\begin{figure}[htb!]
\center
\includegraphics[width=0.9\hsize]{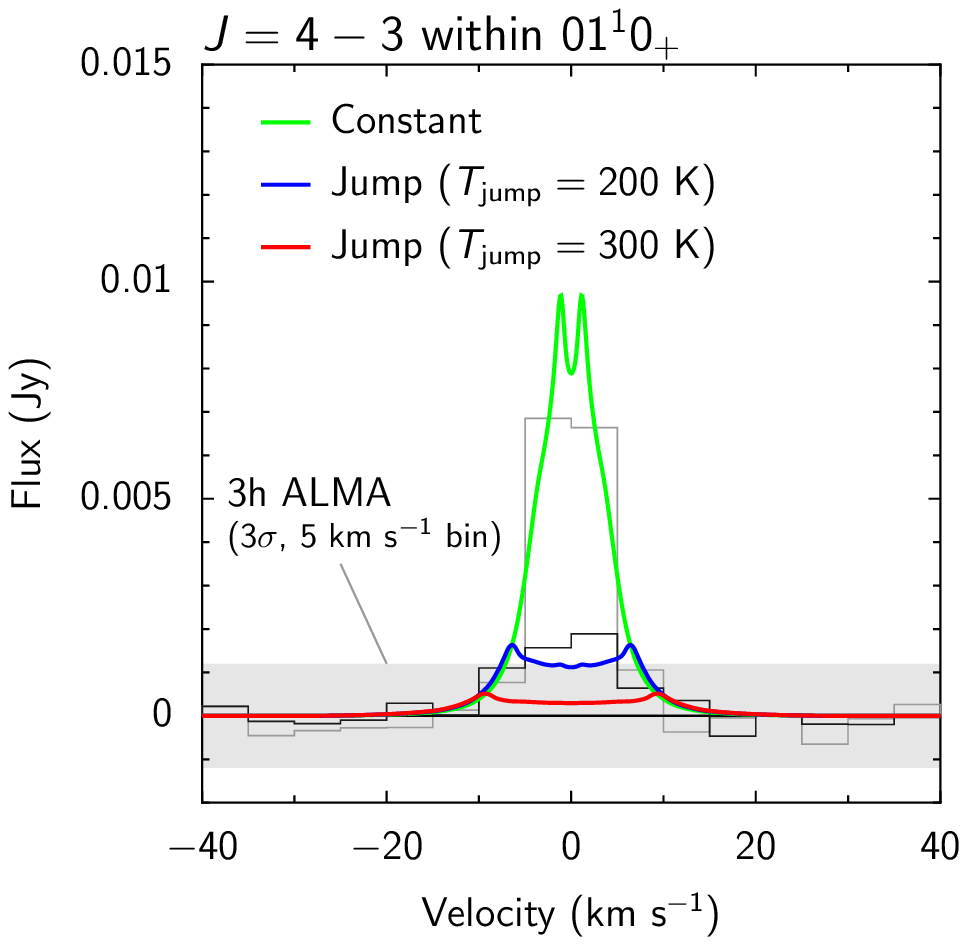}
\caption{Predictions for ALMA. The $J=4-3$ line within the $01^10_+$ band is shown for the models given in Table \ref{tab:flx_constjump} with G/D$=1000$ and constant or jump abundance. The ALMA $3 \sigma$ detection limit (5 km s$^{-1}$ bin) within 3 hours is shown by the gray region. Gray lines represent the constant abundance and jump ($T_{\rm jump}=200$ K) model with noise added.}
\label{fig:plot_alma}
\end{figure}

\begin{figure*}[htb!]
\center
\includegraphics[width=1.0\hsize]{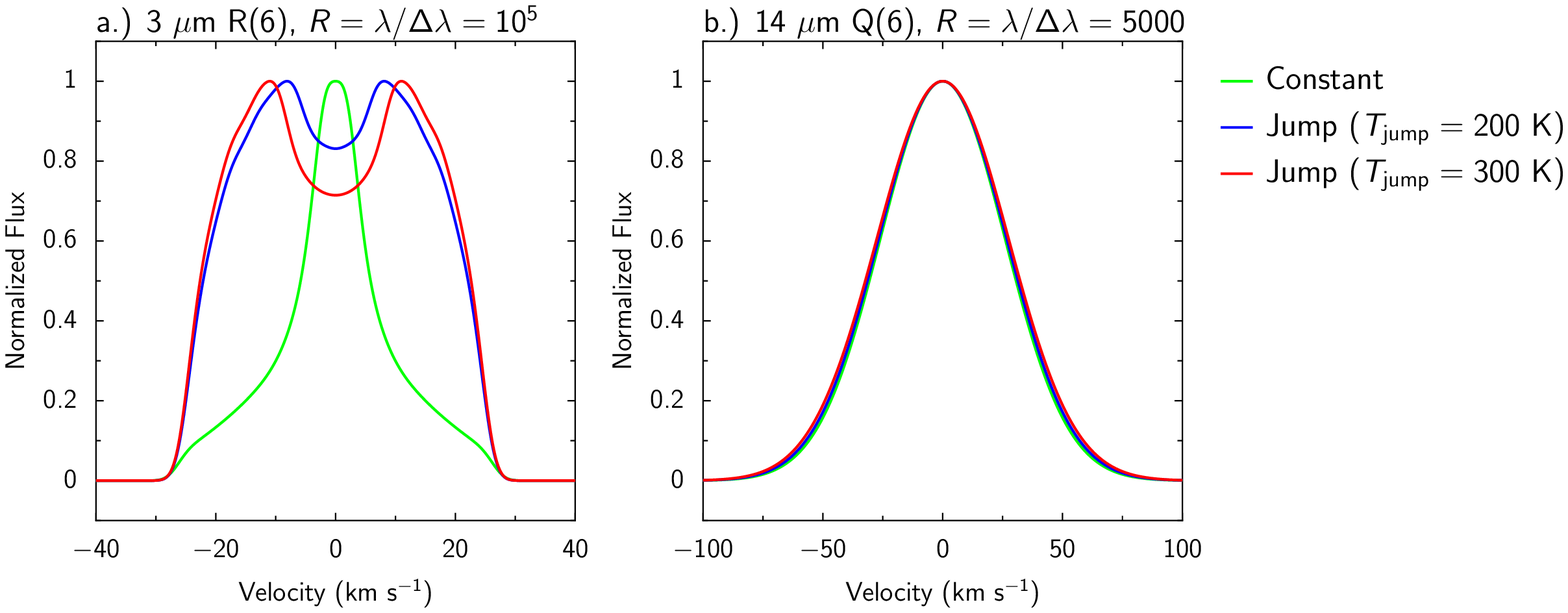}
\caption{Spectrum of the \textbf{a)} R(6) $10^00-00^00$ line at 3 $\mu$m and \textbf{b)} Q(6) $01^10-00^00$ line at 14 $\mu$m of the AS 205 (N) model convolved to the spectral resolution of E-ELT METIS. The models given in Table \ref{tab:flx_constjump} with G/D$=1000$ and constant or jump abundance are shown.}
\label{fig:plot_metis}
\end{figure*}

%
%

\section{HCN ro-vibrational collision rates} \label{sec:app_coll}

The vibrational H$_2$-HCN collisional de-excitation rate coefficients at room temperature (297 K) of three transitions have been measured by \cite{Smith91}. They find for the \emph{total} vibrational de-excitation rate coefficients between an excited state and the ground state\footnote{Note that their level designation has reversed $\nu_1$ and $\nu_3$ with respect to ours, see \cite{Cannon84}. In this section we use the same designation as in the remainder of our work.}
\begin{eqnarray*}
k(\nu_1 \nu_2 \nu_3=100 \rightarrow 000) &=& 4.70 \times 10^{-14} \ \ {\rm cm}^3 {\rm s}^{-1} \\
k(110 \rightarrow 000) &=& 1.28 \times 10^{-11} \ \ {\rm cm}^3 {\rm s}^{-1} \\
k(101 \rightarrow 000) &=& 1.94 \times 10^{-13} \ \ {\rm cm}^3 {\rm s}^{-1} 
\end{eqnarray*}
Thus, the excitation of the stretching modes is much slower than the bending mode. Hence, we will assume that 
\begin{eqnarray*}
k(010 \rightarrow 000) &=& k(110 \rightarrow 000) \\
k(001 \rightarrow 000) &=& k(101 \rightarrow 000) 
\end{eqnarray*}
The temperature dependence of the rate coefficients is not know, but comparison with a triatomic system with similar vibrational energy and reduced mass (CO$_2$-H$_2$, \citealt{Boonman03c} or H$_2$O-H$_2$, \citealt{Faure08}) suggest, that the rate coefficients do not change by more than a factor of a few from 300 to 1000 K. This may be different for low temperatures $T \ll 300$ K, but such cold regions do not affect the results presented in this work. 

Even though strictly applicable only for stretching modes of diatomic molecules, we will approximate the rate coefficients of other transitions which only change either $\nu_i=\nu_1$, $\nu_2$ or $\nu_3$ using the Landau-Teller relationship for transitions between neighboring states (\citealt{Procaccia75} and Eqs. 6 and 8 in \citealt{Chandra01}),
\begin{eqnarray}
k(\nu_i=2 \rightarrow 1) &=& k(\nu_i=1 \rightarrow 0) \times 2 \\
k(\nu_i=2 \rightarrow 0) &=& k(\nu_i=1 \rightarrow 0) \times 2/3 \\
k(\nu_i=0 \rightarrow 1) &=& k(\nu_i=1 \rightarrow 0) \times \exp\left(-h c \omega_i / k T\right)\\
k(\nu_i=1 \rightarrow 2) &=& k(\nu_i=1 \rightarrow 0) \times 2 \, \exp\left(-h c \omega_i / k T\right)\\
k(\nu_i=0 \rightarrow 2) &=& k(\nu_i=1 \rightarrow 0) \times 5/3 \times \exp\left(-h c \omega_i / k T\right)
\end{eqnarray}
Here, $\omega_i$ is the vibrational constant (cm$^{-1}$). In case a transition changes combinations of $\nu_1$, $\nu_2$, and $\nu_3$, it is assumed that the rate coefficient is the maximum of the above rate coefficient changing one mode at the time. 

Table \ref{tab:coll_band} shows the vibrational de-excitation rate coefficients calculated in this way for a temperature of 300 and 500 K.

To obtain a full set of ro-vibrational collisional rate coefficients, we employ the method suggested by \cite{Faure08}: Assuming a decoupling of the rotational and vibrational levels, we can write
\begin{equation}
k(v,J \rightarrow v',J';T)=P_{v,v'}(T) \times k(0,J \rightarrow 0,J';T)
\end{equation}
where 
\begin{equation}
P_{v,v'}(T) = \frac{k(v \rightarrow v') \sum_J g_J \exp\left(-E_{v,J}/k T\right)}
{\sum_J g_J \exp\left(-E_{v,J}/k T\right) \sum_{J'} k(0,J \rightarrow 0, J'; T)}, 
\end{equation}
with the statistical weights $g_i$ of the levels. This procedure ensures that the detailed balance is fulfilled. As \cite{Faure08} we assume that pure rotational rate coefficients within one vibrational level are equal to the ground state rate coefficients.

\begin{figure*}
\center
\sidecaption
\includegraphics[width=0.6\hsize]{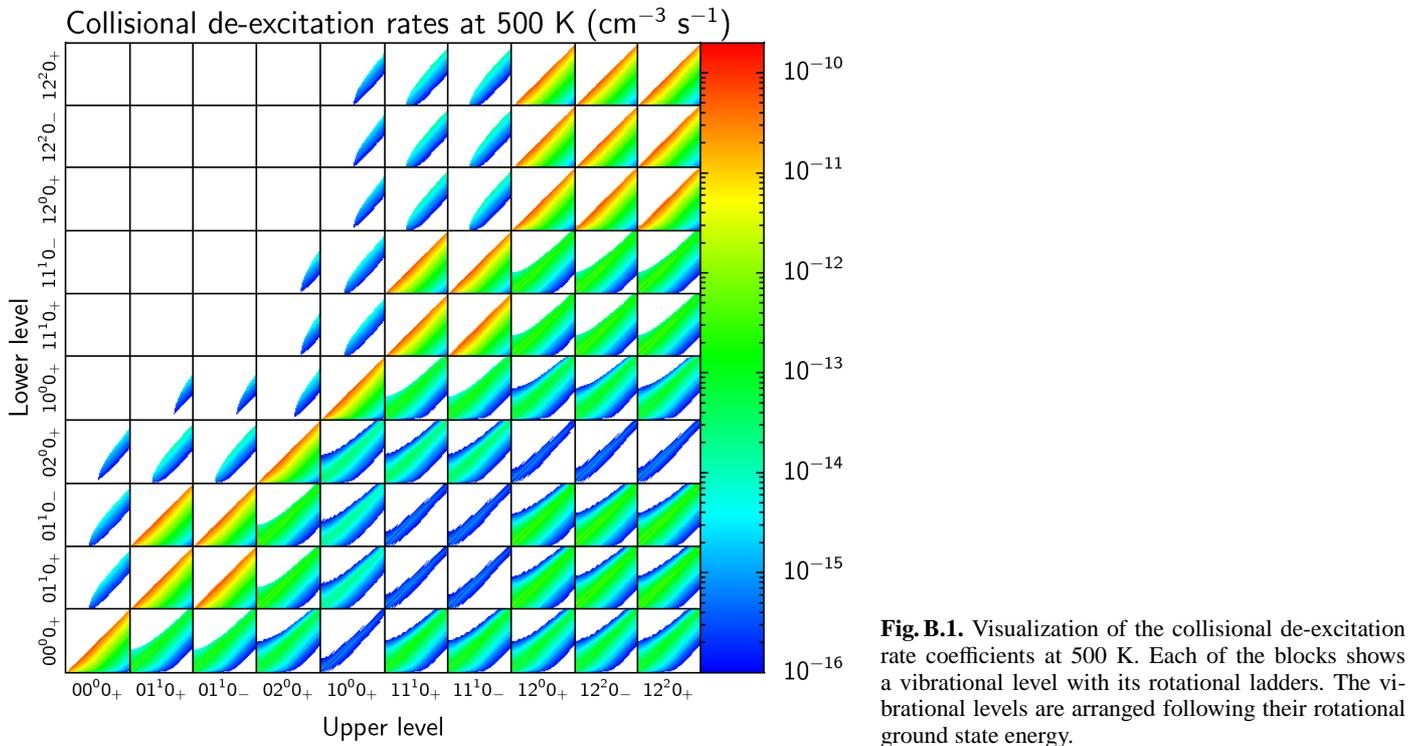}
\caption{Visualization of the collisional de-excitation rate coefficients at 500 K. Each of the blocks shows a vibrational level with its rotational ladders. The vibrational levels are arranged following their rotational ground state energy.}
\label{fig:plot_collrates_500K}
\end{figure*}

The rate coefficients for the pure rotational transitions $k(0,J \rightarrow 0, J';T)$ are taken from \cite{Dumouchel08}. The He-HCN have been scaled by the reduced weight of the H$_2$-HCN system and levels with $J>26$ are extrapolated using the Infinite Order Sudden (IOS) approximation as described in Section 6 of \cite{Schoier05}. The full set of derived de-exciatation rate coefficients is visualized in Figure \ref{fig:plot_collrates_500K} for a temperature of 500 K.

\begin{table*}
\caption{Vibrational de-excitation rate coefficients $k(v \rightarrow v')$ in cm$^3$ s$^{-1}$.}
\label{tab:coll_band}
\centering
\begin{tabular}{cccc}
\hline\hline
Initial state $v$ & Final state $v'$ & $k(v \rightarrow v';{\rm 300 K})$ & $k(v \rightarrow v';{\rm 500 K})$ \\ 
\hline
$01^10$ & $00^00$ & 1.28(-11) & 1.28(-11) \\
$02^00$ & $00^00$ & 8.53(-12) & 8.53(-12) \\
$02^00$ & $01^10$ & 2.56(-11) & 2.56(-11) \\
$10^00$ & $00^00$ & 4.70(-14) & 4.70(-14) \\
$10^00$ & $01^10$ & 3.92(-13) & 1.58(-12) \\
$10^00$ & $02^00$ & 6.53(-13) & 2.63(-12) \\
$11^10$ & $00^00$ & 1.28(-11) & 1.28(-11) \\
$11^10$ & $01^10$ & 4.70(-14) & 4.70(-14) \\
$11^10$ & $02^00$ & 7.83(-13) & 3.16(-12) \\
$11^10$ & $10^00$ & 1.28(-11) & 1.28(-11) \\
$12^00$ & $00^00$ & 8.53(-12) & 8.53(-12) \\
$12^00$ & $01^10$ & 2.56(-11) & 2.56(-11) \\
$12^00$ & $02^00$ & 4.70(-14) & 4.70(-14) \\
$12^00$ & $10^00$ & 8.53(-12) & 8.53(-12) \\
$12^00$ & $11^10$ & 2.56(-11) & 2.56(-11) \\
\hline
\end{tabular}
\tablefoot{$a(b)$ means $a \times 10^{b}$}
\end{table*}

\section{Line-to-continuum ratio} \label{sec:app_lcratio}

\begin{figure*}
\center
\includegraphics[width=0.84\hsize]{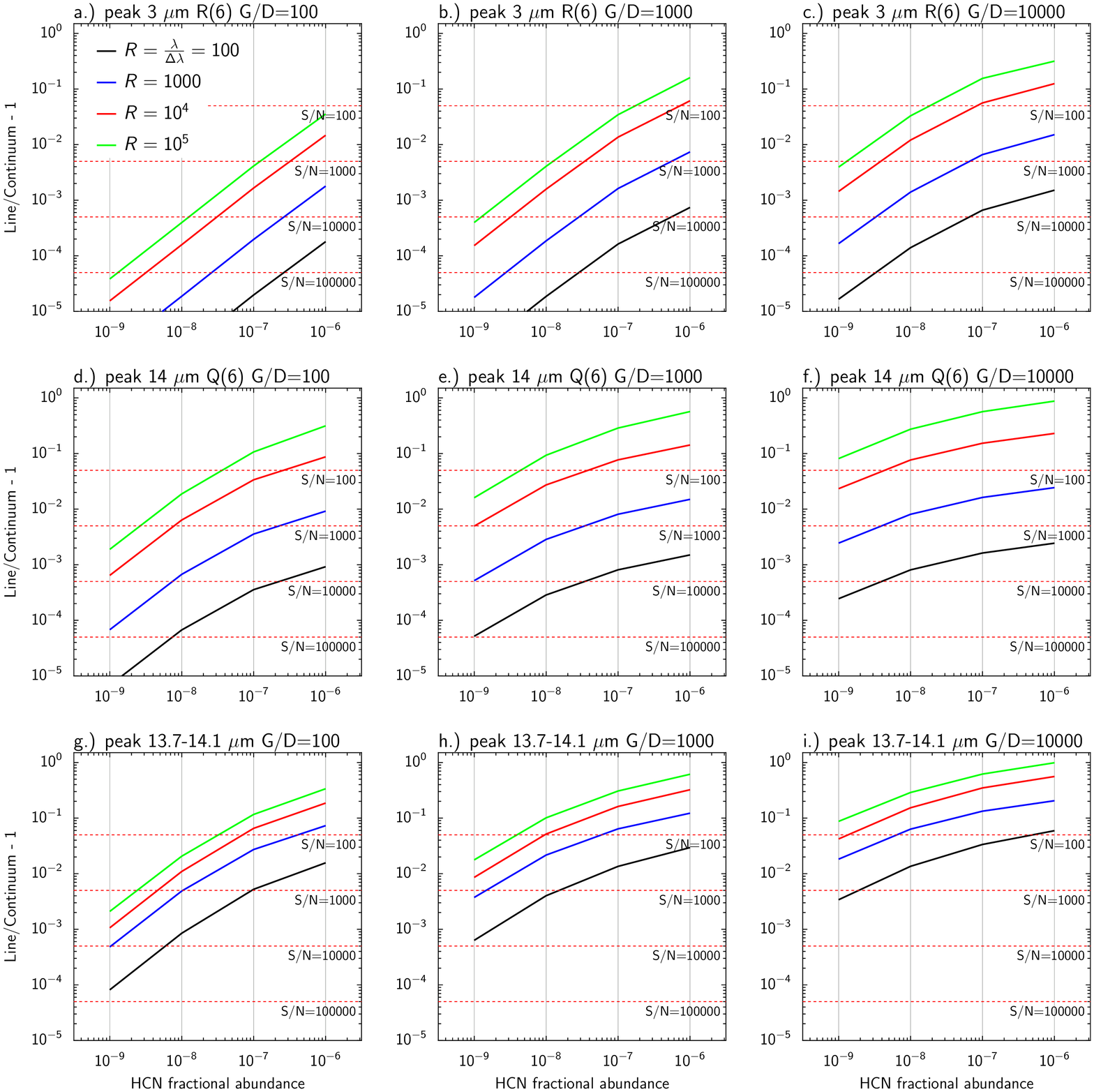}
\caption{The peak line-to-continuum ratio of the \textbf{a)-c)} R(6) $10^00-00^00$ line at 3 $\mu$m, \textbf{d)-f)} Q(6) $01^10-00^00$ line at 14 $\mu$m, and \textbf{g)-i)} between 13.7 and 14.1 $\mu$m. The non-LTE models with a constant abundance and a gas-to-dust (G/D) ratio of 100, 1000 and 100000 are shown (Figure \ref{fig:plot_diskflx_ltenlte}), convolved to a spectral resolving power $R=\lambda/\Delta\lambda$ between 100 and $10^5$. Horizontal red dashed lines show the required signal-to-noise (S/N) in the continuum to detect a line at a $5\sigma$ level.}
\label{fig:plot_lcratio}
\end{figure*}

The line-to-continuum ratio can be the limiting factor to detect lines. For example the CRIRES detections by \citet{Mandell12} with line-to-continuum ratios of $\sim 1$ \% in the 3 $\mu$m lines required a signal-to-noise in the continuum of several 100. Figure \ref{fig:plot_lcratio} illustrates the line-to-continuum ratio for the non-LTE models with a constant abundance and different gas-to-dust ratios. The peak line-to-continuum ratio for different spectral resolving power ($R=\lambda/ \Delta \lambda$) is shown for the R(6) $10^00-00^00$ line at 3 $\mu$m, the Q(6) $01^10-00^00$ line at 14 $\mu$m, and between 13.7 to 14.1 $\mu$m (mostly Q-branch of the $01^10-00^00$ band). The required signal-to-noise ratio in the continuum to detect a line at a $5\sigma$ level is shown in the figure.

The line-to-continuum ratio increases with the spectral resolution. The increase is strongest at low spectral resolution in the single lines. The 13.7-14.1 $\mu$m range, where several lines contribute to the peak line-to-continuum ratio, has a higher peak line-to-continuum ratio at low $R$. To detect the 3 $\mu$m lines, high resolution spectroscopy ($R=10^5$) should be used, since these lines have lower line-to-continuum ratios compared to the 14 $\mu$m lines. Peak line-to-continuum ratios and line fluxes show the same degeneracy between the abundance and gas-to-dust ratio (Section \ref{sec:lte_non_lte}). A factor of ten higher abundance, but by the same factor lower gas-to-dust ratio yields approximately the same line-to-continuum ratio. For MIRI with a spectral resolving power of $R=3000$, a gas-to-dust ratio of 1000 and an abundance of $3 \times 10^{-8}$ (Table \ref{tab:flx_constjump}), a signal-to-noise of 1000 in the continuum will allow detecting lines down to 5 \% of the peak of the Q-branch, i.e., P- and R-branch lines from high-$J$ levels.

\end{appendix}

\end{document}